\documentclass[useAMS,usenatbib, aastex]{mn2e}

\usepackage{float}
\usepackage{hyperref} % For hyperlinks in the PDF
\usepackage{graphicx}
\usepackage{amssymb}
\usepackage{epstopdf} 
\usepackage{natbib}
\usepackage{aas_macros}

\title[The Sagittarius Dwarf Irregular]{Solo Dwarfs I: Survey introduction and first results for the Sagittarius Dwarf Irregular Galaxy}
\author[Higgs et al.]{C. R. Higgs$^{1,2}$\thanks{E-mail: higgs@uvic.ca}, A.W. McConnachie$^{2}$, M. Irwin$^{3}$, N. F. Bate$^{3}$, G. F. Lewis$^{4}$,\newauthor M. G. Walker$^5$, P. C{\^o}t{\'e}$^{2}$, K. Venn$^{1}$, G. Battaglia$^{6,7}$\\
$^{1}$University of Victoria, Victoria, B.C., Canada\\
$^{2}$NRC Herzberg Institute of Astrophysics, 5071 West Saanich Road, Victoria, B.C., V9E 2E7, Canada\\
${^3}$Institute of Astronomy,  Madingley  Road,  Cambridge, CB3  0HA,  U.K.\\
$^4$Sydney Institute  of Astronomy, School  of Physics, A28, The University of  Sydney, NSW 2006, Australia\\
$^5$Department of Physics, Carnegie Mellon University, 5000 Forbes Ave., Pittsburgh, PA 15213\\
$^6$Instituto de Astrofisica de Canarias, calle via Lactea s/n, 38205 San Cristobal de La Laguna (Tenerife), Spain\\
$^7$Universidad de La Laguna, Dpto. Astrofisica, E-38206 La Laguna, Tenerife, Spain\\}

\begin{document}

\date{\today}

\pagerange{\pageref{firstpage}--\pageref{lastpage}} \pubyear{2014}

\maketitle

\label{firstpage}

\begin{abstract}

We introduce  the {\it So}litary {\it Lo}cal Dwarfs Survey ({\it Solo}), a wide field photometric study targeting every isolated dwarf galaxy within 3 Mpc of the Milky Way. {\it Solo} is based on $(u)gi$ multi-band imaging from CFHT/MegaCam for northern targets, and Magellan/Megacam for southern targets. All galaxies fainter than $M_V \simeq -18$ situated beyond the nominal virial radius of the Milky Way and M31 ($\gtrsim$ 300 kpc) are included in this volume-limited sample, for a total of 42 targets. In addition to reviewing the survey goals and strategy, we present results for the Sagittarius Dwarf Irregular Galaxy (Sag DIG), one of the most isolated, low mass galaxies, located at the edge of the Local Group. We analyze its resolved stellar populations and their spatial distributions. We provide updated estimates of its central surface brightness and integrated luminosity, and trace its surface brightness profile to a level fainter than 30 mag./sq.arcsec. Sag DIG is well described by a highly elliptical (disk-like) system following a single component Sersic model. However, a low-level distortion is present at the outer edges of the galaxy that, were Sag DIG not so isolated, would likely be attributed to some kind of previous tidal interaction. Further, we find evidence of an extremely low level, extended distribution of stars beyond $\sim 5$ arcmins ($> 1.5 $ kpc) that suggests Sag DIG may be embedded in a very low density stellar halo. We compare the stellar and HI structures of Sag DIG,  and discuss results for this galaxy in relation  to other isolated, dwarf irregular galaxies in the Local Group.
\end{abstract}


\section{Introduction}

Dwarf galaxies form a surprisingly large morphological menagerie, including (but probably not limited to): blue-compact, elliptical, irregular, low surface brightness, Magellanic-type, spheroidal, spiral, transition-type, ultra-compact, and ultra-faint. 
The recently discovered ultra-diffuse dwarfs (\citealt{Vandokkum2015}) highlight their variety. The drivers of this diversity are the focus of much of the literature on these objects. While certainly governed by the same physics as their more massive counterparts, dwarf galaxies are generally expected to be more sensitive to the processes that drive galaxy formation and evolution. 
Both internally driven processes, like star formation and stellar feedback, and externally driven processes, like ram pressure or tidal stripping, can greatly impact the structure and morphology of these small systems.

There have been a large number of recent discoveries of dwarf galaxies in the Local Group, particularly in the vicinity of the Milky Way and M31, often with exceptionally low stellar mass. With the addition of these very low mass systems, \cite{Tolstoy2009} discuss the new found difficulty in differentiating faint dwarf galaxies from stellar clusters (see also \citealt{Willman2012}). A significant dark matter component is generally considered to be the major distinguishing factor, usually identified through a high stellar velocity dispersion based on radial velocity measurements of many stars. However, in some cases the data is not conclusive, due either to only a few stars being accessible for measurement (e.g., \citealt{Simon2007, Kirby2015}) or uncertainties on the conversion from a radial velocity dispersion to a dynamical mass (e.g., \citealt{Mcconnachie2010a} and references therein). In some cases without reliable velocity measurements, a spread in iron abundance can be used as indirect evidence of dark matter (with the implication that a deep gravitational potential is required to retain the enrichment products of Type II supernovae; e.g., see \citealt{Kirby2015} for a recent example). 

At the bright end of the dwarf luminosity distribution, \cite{Kormendy1985} famously demonstrated that ``dwarf" galaxies appear to define a scaling relation between surface brightness and magnitude that is nearly orthogonal to ``giant" galaxies. Recently, however, growing evidence suggests that dwarfs and giants define a single, continuous but non-homologous relation in this parameter space (\citealt{Ferrarese2012}). Mindful of these considerations, we adopt the generally-accepted definition of a dwarf galaxy as having an absolute magnitude of $M_V > -18$ (e.g., \citealt{Grebel2003, McConnachie2012}).

The low stellar mass and low surface brightness of dwarf galaxies mean that it is often not possible to examine their detailed properties out to a significant redshift. Indeed, for the lowest stellar mass systems such as Segue I (with a stellar mass of only a few thousand stellar masses; \citealt{Belokurov2007}), it is unlikely that they would have been discovered even if they were located at just $>0.5\,$Mpc away (\citealt{Koposov2008}). However, by resolving individual stars and searching for stellar over-densities rather than relying on detection of integrated light, extremely faint galaxies can be identified in the vicinity of the Local Group. The ability to resolve individual stars also opens up a host of related studies, allowing us to gain insight into structural properties, star formation histories, and chemical evolution that eludes studies of more distant galaxies (see, for example, reviews by \citealt{Mateo1998a, Tolstoy2009, McConnachie2012})

Due to the proximity bias described above for dwarf galaxies,  the most detailed studies of dwarfs are heavily biased towards the Milky Way satellites (and, to a lesser extent, the M31 satellites).
  With some notable exceptions, these dwarfs are generally devoid of gas and show no sign of ongoing star formation. Therefore, most are classified as dwarf spheroidal (or dwarf elliptical if they are brighter than $M_V \simeq -15$). In contrast, it has long been known (\citealt{Einasto1974}) that more isolated dwarf galaxies in the Local Group are frequently found to be gas rich and have young stellar populations, indicative of ongoing or very recent star formation (and are referred to as dwarf irregular galaxies). The position--morphology relation in the Local Group is seen in other groups as well (\citealt{Bouchard2009}, \citealt{Gallart2015}) and suggests a close link between the nearby presence of a large galaxy and the evolution of dwarf galaxies. A key study by \cite{Geha2012} using dwarf galaxies identified in the SDSS has shown that the influence of a large host galaxy can be detected in the average properties of dwarf galaxies out to a distance of some 1.5 Mpc from the host.

It has been proposed that ram pressure stripping by a hot gaseous halo surrounding a large galaxy, perhaps working in concert with tidal effects such as stripping and stirring, could remove the gas from a dwarf-irregular type galaxy and transform it to a dwarf spheroidal galaxy (in particular, see the models by \citealt{Mayer2006} and collaborators). Certainly, there are some well known examples of tidal stripping, such as the Sagittarius dwarf spheroidal galaxy (\citealt{Ibata1994}) which has an enormous stellar stream spanning the entire sky that reveals its interaction history with the Milky Way (\citealt{Ibata2001b, Majewski2003} and references therein). There are  also numerous examples of galaxies undergoing ram-pressure stripping due to interactions in cluster-like 
environments (e.g.,  NGC4522 in the Virgo cluster, \citealt{Vollmer2000}; Peg DIG in the Local Group, \citealt{McConnachie2007b}; see also \citealt{Boselli2014}). But the proximity of a large galaxy can introduce more subtle effects in the properties of the smaller satellite. For example, interactions could lead to disk instabilities in the satellite and produce warping. In the right circumstances, rotationally supported systems could be transformed into pressure supported systems (\citealt{Mayer2006}). Interactions may also trigger star formation if there is gas still present in the dwarf (particularly at pericenter), but conversely could also prevent the dwarf from accreting new gas to replenish its supply, thus leading to the eventual termination of star formation (so called ``strangulation", see \cite{Kawata2008}, among others). 

The net result of a large body of work is a demonstration that the morphological, dynamical, star forming and  chemical evolutionary status of dwarf galaxies can be severely affected by proximity to large galaxies. However, many of the ``symptoms" of interactions are not unique and indeed may also be explained via secular processes. For example, gas could be lost from a low mass galaxy due to feedback from supernovae (\citealt{Dekel1986, Dekel2003}), without requiring an external actor. 

With these considerations in mind, we have constructed the {\it Solo} dataset. Here, our intent is to provide homogeneous, high quality, wide field optical characterization and study of the closest, isolated dwarf galaxies whose resolved stellar content is accessible, at least in part, using current ground based facilities. This sample therefore constitutes all known isolated dwarf galaxies for which we can expect to obtain, now or in the near future, the most detailed perspectives on their structures, dynamics, star formation and chemical evolution that is of a comparable nature and quality to the information accessible for the Milky Way and M31 satellite galaxies. When viewed as a population, the {\it Solo} dwarfs are potentially a powerful benchmark for studies of the interplay between secular and environment-driven galaxy evolution at the low mass end. 

Section 2 describes the selection criteria and reduction techniques for the {\it Solo} dataset, and we review in detail the literature pertaining to the Sagittarius dwarf irregular (Sag DIG), the first galaxy analyzed as part of this program and the focus of the remainder of this paper. Section 3 presents a colour magnitude analysis of Sag DIG, including distance and metallicity estimates, stellar population characterization and analysis of radial gradients. Section 4 presents a study of the structural properties of Sag DIG to very low surface brightness. In Section 5,  we discuss our results in the context of previous results for this galaxy and in comparison to other Local Group dwarf irregular galaxies. Finally, in Section 6, we summarize.

\section{Preliminaries}

\subsection{The Solo Dwarfs}

The {\it Solo} dwarf galaxies are selected from the compilation of dwarf galaxies presented in \cite{McConnachie2012}. Briefly, \cite{McConnachie2012} presents all dwarf galaxies that have distance estimates based on measurements of resolved stellar populations (i.e., Cepheids, RR Lyrae, tip of the red giant branch, main sequence fitting) that place them within 3 Mpc of the Sun. A regularly updated version of this catalog is available online\footnote{http://www.astro.uvic.ca/$\sim$alan/Nearby\_Dwarf\_Database.html}. The distance threshold of 3 Mpc corresponds to the approximate distance to the next nearest galaxy groups to the Local Group (\citealt{Karachentsev2005}). More than 100 galaxies are known within 3 Mpc,  with several discovered since the original publication of \citealt{McConnachie2012}. The majority of these new discoveries are Milky Way satellites. However, two new isolated galaxies - KK258 and KKs3 - have also been ``discovered" in the neighbourhood of the Local Group, as updated distances from HST have shown them to be closer than previously estimated (\citealt{Karachentsev2014, Karachentsev2015}). This brings the total number of galaxies within 3 Mpc lying beyond the nominal virial radii of the Milky Way and M31 (adopted to be 300 kpc, e.g., \citealt{Klypin2002}) to 42. 

\begin{figure}
\begin{center}
\includegraphics[width=\linewidth]{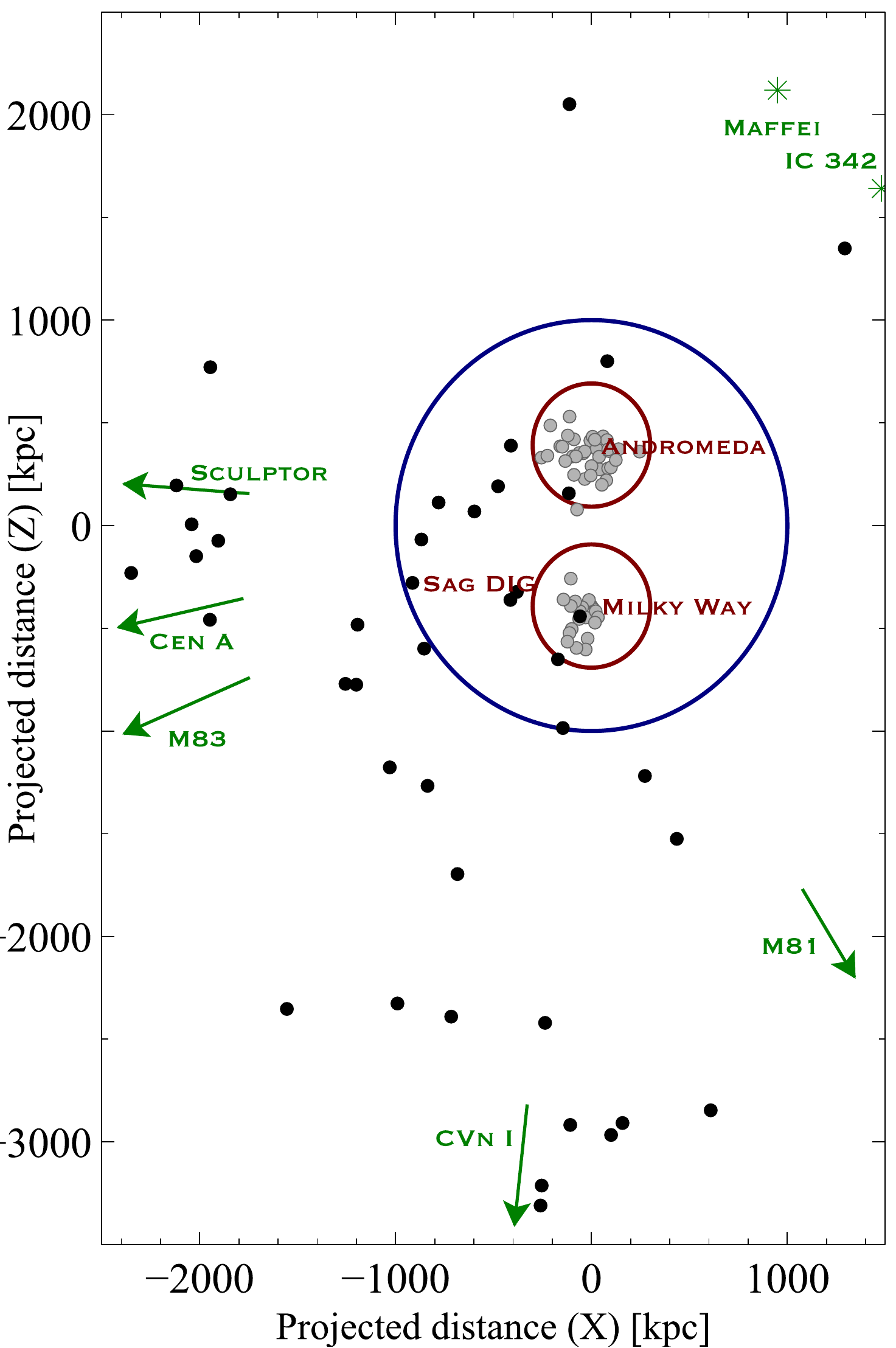}
\caption{Projected distribution of dwarf galaxies within 3 Mpc, with M31 (Andromeda) and the Milky Way labelled. The projection is centered on the point halfway between the Milky Way and M31, with the y axis aligned with the M31 - Milky Way direction. Black points indicate isolated dwarfs in the {\it Solo} sample. Grey points are M31 and Milky Way satellites (defined to lie within 300 kpc of their host, indicated by the red circles).  Green symbols indicate the positions or directions of the nearest galaxy groups to the Milky Way. The blue circle indicates the zero velocity surface (as determined by \protect\cite{McConnachie2012}) of the Local Group.}
\label{sample}
\end{center}
\end{figure}

Figure \ref{sample} shows a projection of all the dwarf galaxies within 3 Mpc using the positions and distances given in \cite{McConnachie2012}. 
The isolated {\it Solo} sample (black points) is shown along with the satellites of M31 and the Milky Way (grey points).
For the purposes of this work, we consider any dwarf located within 300 kpc of either the Milky Way or M31 as a satellite, and any dwarf located beyond this threshold as isolated. All dwarfs meeting this latter criterion are considered part of the  {\it Solo} survey. We recognize that this (semi-arbitrary) cut will not uniquely select galaxies that have never interacted with either the Milky Way or M31 (for example, see the study by \citealt{Geha2012}). It may include some dwarfs that are on very long period orbits, or ``backsplash" galaxies that are at very large distances but which have had a previous pericentric passage with one of the two large bodies (e.g., \citealt{Gill2005}). Numerical simulations of Local Group environments are a particularly useful comparison to observational datasets in understanding the orbital diversity of the dwarf population (e.g., \citealt{Barber2014, Garrisonkimmel2014}). Timing arguments suggest that most galaxies located near or beyond the periphery of the Local Group will not have had time to have had a close interaction at any point in their history (\citealt{McConnachie2012}). Table \ref{sampletable} lists all the dwarf galaxies in the {\it Solo} survey, along with their positions, current distance estimates, and relevant observational details.

All the galaxies listed in Table \ref{sampletable} are close enough to resolve their brighter stellar populations (at least in their outskirts) from the ground. Many of the closest targets in this list are reasonably well studied, particularly the members of the Local Group. However, a systematic, modern survey including more distant dwarfs is lacking.  Arguably the most systematic study of a subset of these galaxies is \cite{Massey2006}, who surveyed ten star forming galaxies to determine certain properties relating to star formation activity. In addition, the Local Cosmology from Isolated Dwarfs (LCID) survey studied six nearby isolated dwarf galaxies in the Local Group using deep HST imaging (see \citealt{Gallart2015} and references within). This imaging reaches the main sequence turn off for these dwarfs, and has provided some of the most detailed insights into the star formation histories and stellar content of these galaxies available (e.g., \citealt{Monelli2010a, Monelli2010, Hidalgo2011, Hidalgo2013, Skillman2014, Gallart2015}). Many of the more distant dwarfs listed in Table \ref{sampletable} are relatively poorly studied in the era of modern wide field CCD studies. An examination of the data tables in \cite{McConnachie2012} reveal that many of the dwarfs' basic properties are derived from the Third Reference Catalog survey by \cite{Devaucouleurs1991}, and a homogeneous wide field study of the entire sample has been conducted recently. Of course, there are notable programs that have focused on other aspects of the properties of these systems. For example,  \cite{Dalcanton2009} have studied many nearby targets using HST/ACS and WFPC2. HST is ideal for studying the resolved stars in the centers of the galaxies, but its limited field of view makes it less ideal for wide-field structural and stellar content studies. HST has been used for structural studies of these galaxies. For example, \cite{Hidalgo2009} uses two HST fields to look for gradients in stellar populations in the Phoenix dwarf. However, the total luminosity of this galaxy (\citealt{Vanderydt1991}) is still based on imaging that covers only part of the galaxy, and no direct measurement has ever made of the integrated light of this galaxy.  Overall, the lack of a systematic, homogeneous wide field study of the nearby dwarf population means that much of their basic data may still be uncertain. It is this important niche that {\it Solo} is designed to filled, while being complementary to the ground and space based studies that have been undertaken of individual targets.

\begin{table*}
 \centering
 \begin{minipage}{140mm}
  \caption{The {\it Solo} dwarfs. ``C" indicates observations with CFHT/MegaCam, and ``M" indicates observations with Magellan/Megacam. Distance estimates are from the updated compilation by  \protect\cite{McConnachie2012}.  $E(B-V)$ estimates are from the \protect\cite{Schlafly2011} dust maps along the line of sight to the center of the galaxy. } \label{sampletable}
  \begin{tabular}{@{}llllcccl@{}}
  \hline
  \hline
   Name     &      RA      &  Dec. & Distance &$E(B-V)$ & Telescope & Filters & Notes\\
    &&& [kpc] &[Mags.]&&\\
\hline
 \hline
\\

{\it Complete data}
\\
WLM            &    $00^{h}01^{m}58.2^{s}$     &   $-15^{o}27{'}39"$   &  $933  \pm 34$ & 0.04 & M & $g i$ & 2014B\\ 
               & & & &                                                                          & C & $g$  & 2013B \\
AndXVIII       &    $00^{h}02^{m}14.5^{s}$     &   $+45^{o}05{'}20"$   & $1355  \pm 81$ & 0.10 & C & $g i$ & PAndAS data \\ 
ESO410-G005    &    $00^{h}15^{m}31.6^{s}$     &   $-32^{o}10{'}48"$   & $1923  \pm 35$ & 0.01& M & $u g i$ & 2012B\\ 
Cetus          &    $00^{h}26^{m}11.0^{s}$     &   $-11^{o}02{'}40"$   & $755   \pm 24$ & 0.03 & M & $g i$ & 2014B\\
ESO294-G010    &    $00^{h}26^{m}33.4^{s}$     &   $-41^{o}51{'}19"$   & $2032  \pm 37$ & 0.01 & M & g i & 2014B \\
IC1613         &    $01^{h}04^{m}47.8^{s}$     &   $+02^{o}07{'}04"$   & $755   \pm 42$ & 0.03 & M & $u g i$ & 2012B\\
HIZSS3A(B)     &    $07^{h}00^{m}29.3^{s}$     &   $-04^{o}12{'}30"$   & $1675  \pm 108$& 0.69 & M & $u g i$  & 2012A\\
               & & & &                                                                          & C & $g$  & 2012A \\
NGC3109        &    $10^{h}03^{m}06.9^{s}$     &   $-26^{o}09{'}35"$   & $1300  \pm 48$ & 0.07 & M & $u g i$  & 2012A \\
Antlia         &    $10^{h}04^{m}04.1^{s}$     &   $-27^{o}19{'}52"$   & $1349  \pm 62$ & 0.08 & M & $g i$ & 2015A, IMACS \\
SextansA       &    $10^{h}11^{m}00.8^{s}$     &   $-04^{o}41{'}34"$   & $1432  \pm 53$ & 0.05 & M  & $u g i$ & 2012A \\
LeoP           &    $10^{h}21^{m}45.1^{s}$     &   $+18^{o}05{'}17"$   & $1620  \pm 150$ & 0.03 & C & $g i$ & 2014A \\
IC3104         &    $12^{h}18^{m}46.0^{s}$     &   $-79^{o}43{'}34"$   & $2270  \pm 188$ & 0.30 & M & $u g i$  & 2012A \\
GR8            &    $12^{h}58^{m}40.4^{s}$     &   $+14^{o}13{'}03"$   & $2178  \pm 120$ & 0.03 & M & $g i$ & 2015A, IMACS \\
KKH86          &    $13^{h}54^{m}33.5^{s}$     &   $+04^{o}14{'}35"$   & $2590  \pm 190$ & 0.03 & M & $u g i$ & 2012A\\
DDO190         &    $14^{h}24^{m}43.4^{s}$     &   $+44^{o}31{'}33"$   & $2790  \pm 93$ & 0.01  & C & $g i$  & 2014A\\
KKR25          &    $16^{h}13^{m}48.0^{s}$     &   $+54^{o}22{'}16"$   & $1905  \pm 61$ & 0.01  & C & $g i$ & 2014A \\
Sag DIG         &    $19^{h}29^{m}59.0^{s}$     &   $-17^{o}40{'}41"$   & $1067  \pm 88$ & 0.12  & C & $u g i$ & 2012B, 2013A \\
NGC6822        &    $19^{h}44^{m}56.6^{s}$     &   $-14^{o}47{'}21"$   & $459   \pm 17 $& 0.19  & C & $u g i$  & 2013A  \\
Phoenix        &    $19^{h}44^{m}56.6^{s}$     &   $-14^{o}47{'}21"$   & $415   \pm 19 $& 0.02  & M  & $u g i$ & 2012B\\
DDO210         &    $20^{h}46^{m}51.8^{s}$     &   $-12^{o}50{'}53"$   & $1072  \pm 39$ & 0.05  & M & $u g i$ & 2012B\\
               & & & &                                                                           & C & $u g i$  & 2013A \\
IC5152         &    $22^{h}02^{m}41.5^{s}$     &   $-51^{o}17{'}47"$   & $1950  \pm 45$ & 0.03  & M & $u g i$ & 2012B\\
AndXXVIII      &    $22^{h}32^{m}41.2^{s}$     &   $+31^{o}12{'}58"$   & $661^{+152}_{-61}$ & 0.09 &  C & $u g i$ & 2012A, 2013A \\
Tucana         &    $22^{h}41^{m}49.6^{s}$     &   $-64^{o}25{'}10"$   & $887   \pm 49$ & 0.03     & M & $u g i$  & 2012B\\
UKS2323-326    &    $23^{h}26^{m}27.5^{s}$     &   $-32^{o}23{'}20"$   & $2208  \pm 92$ & 0.01     & M & $u g i$  & 2012B \\
Peg DIG         &    $23^{h}28^{m}36.3^{s}$     &   $+14^{o}44{'}35"$   & $920   \pm 30$ & 0.07     & C & $u g i$ & 2012B, 2013A  \\
KKH98          &    $23^{h}45^{m}34.0^{s}$     &   $+38^{o}43{'}04"$   & $2523  \pm 105$ & 0.12    & C & $u g i$ & 2013A  \\

\\ 
{\it Incomplete data}
\\
UGCA86       	       &    $03^{h}59^{m}48.3^{s}$     &   $+67^{o}08{'}19"$   & $2960  \pm 232$& 0.65 & C & $g$ & 2012B \\
DDO99          &    $11^{h}50^{m}53.0^{s}$     &   $+38^{o}52{'}49"$   & $2590  \pm 167$ & 0.03 & C & $i$  & 2014A \\
DDO113                   &    $12^{h}14^{m}57.9^{s}$     &   $+36^{o}13{'}08"$    & $2950  \pm 82 $& 0.02 & C & $g$  & 2014A\\
UGC8508                 &    $13^{h}30^{m}44.4^{s}$     &   $+54^{o}54{'}36"$    & $2580  \pm 36$ & 0.02& C & $g$ & 2014A \\
KKR3              	       &    $14^{h}07^{m}10.5^{s}$     &   $+35^{o}03{'}37"$    & $2188  \pm 121$ & 0.01 &  C & $g$  & 2014A \\
UGC9128                 &    $14^{h}25^{m}56.5^{s}$     &   $+23^{o}03{'}19"$    & $2291  \pm 42$ & 0.02 & C & $g$ & 2014A\\

\\
{\it Data pending}
\\ 

KKs3                       &     $02^{h}24^{m}44.4^{s}$     &    $-73^{o}30{'}51{"}$  & $2120 \pm 70$ & 0.05 &  &&\\ 
Perseus     	       &    $03^{h}01^{m}22.8^{s}$     &   $+40^{o}59{'}17"$   &  $785  \pm 65$ & 0.12 &  &\\ 
UGC4879     	       &    $09^{h}16^{m}02.2^{s}$     &   $+52^{o}50{'}24"$   & $1361  \pm 25$ & 0.02 &    \\
LeoT        		       &    $09^{h}34^{m}53.4^{s}$     &   $+17^{o}03{'}05"$   &  $417  \pm 19$ & 0.03 &  \\
LeoA       		       &    $09^{h}59^{m}26.5^{s}$     &   $+30^{o}44{'}47"$   & $798  \pm44 $& 0.02 & \\
SextansB     	       &    $10^{h}00^{m}00.1^{s}$     &   $+05^{o}19{'}56"$   & $1426  \pm 20$ & 0.03 &  \\
NGC4163                 &    $12^{h}12^{m}09.1^{s}$     &   $+36^{o}10{'}09"$    & $2860  \pm 39$ & 0.02 &  \\ 
DDO125   	       &    $12^{h}27^{m}40.9^{s}$     &   $+43^{o}29{'}44"$    & $2580  \pm 59$ & 0.02 &   \\
IC4662                     &     $17^{h}47^{m}08.8^{s}$     &   $-64^{o}38{'}30{"}$    & $2440 \pm 191$ & 0.07&&& \\ 
KK258                      &     $22^{h}40^{m}43.9^{s}$     &   $-30^{o}47{'}59{"}$   & $2230 \pm 50$ & 0.01&&&\\

\hline
\hline
\end{tabular}
\end{minipage}
\end{table*}

\subsection{Data acquisition and processing}

The {\it Solo} dwarfs have been observed with CFHT/MegaCam for mostly (but not exclusively) northern targets, and Magellan/Megacam for most of the southern targets, detailed in Table \ref{sampletable}. CFHT/MegaCam is an array of 40 individual 2048 $\times$ 4612 pixel CCDs arranged in a $9 (11) \times 4$ grid (where the middle 2 rows have 11 CCDs, and the outer 2 rows have 9 CCDs) with a pixel scale of 0.187 arcsecs/pixel, mounted at the prime focus of the 3.6m CFHT on the summit of Mauna Kea. Prior to the 2015A semester, only the central 36 CCDs were in use for science observations, resulting in a rectangular, $0.96 \times 0.94$ degree field of view. Magellan/Megacam is an array of 36 individual 2048 $\times$ 4608 pixel CCDs arranged in a $9  \times 4$ grid with a pixel scale of 0.08 arcsecs/pixel, mounted at the Cassegrain focus of the 6.5m Clay telescope at the Las Campanas Observatory. The resulting field of view is $25 \times 25$ arcmins. In general, we usually observed a single field with CFHT/MegaCam centered on the target, whereas we tried to observe multiple (usually 4) fields to tile the target with Magellan/Megacam to compensate for the smaller field of view. We note that it is a convenient coincidence that the difference in apertures between the telescopes roughly compensates for the difference in field of view between the cameras. Observations were primarily made in the $g$ and $i$ filters in both hemispheres, with some targets also observed in $u$ with lower priority. Some equatorial targets were observed with both telescopes in order to provide transformations between the CFHT and Magellan photometric systems and to ensure a uniform calibration of the dataset, as indicated in Table \ref{sampletable}. 

CFHT/MegaCam is a queue scheduled instrument, and therefore our observing conditions and strategy are very uniform. Exposures are 1800s in both $g$ and $i$, split as four dithered subexposures of 450s. The uniform exposure time adopted for targets at considerably different distances results in varying depths of resolved star analysis. Considerably fainter stellar populations  will be reached for the closest galaxies than for the most distant galaxies. This approach is optimal given the very large amount of time that is otherwise required to push to fainter stellar populations for the most distant targets, where crowding limits the regions in which we could conduct the resolved star analysis anyway. Magellan/MegaCam is a classically-scheduled instrument, and as such our targets were observed over a broader range of conditions and with some variation in observing time. Generally, we observe in dark time and typically have better than 1 arcsec image quality. We observe for $\sim 450$s in $g$ and $\sim900$s in i per field, usually with $\sim 4$ fields per target. For those targets that have $u$ band data, our exposures were generally of order $360s$.

Our data processing is generally the same for each galaxy, with some slight modifications depending on any individual peculiarities of the target, pointing or conditions.
For Sag DIG CFHT/MegaCam data analyzed here, the image processing steps were similar to those followed by \cite{Richardson2011}.  Data were preprocessed by the Elixir system at CFHT, including de-biasing, flat-fielding and fringe-correcting the $i$ band data in 
addition to determining the photometric zero-points. The data was then
transferred to Cambridge Astronomical Survey Unit where the overscan region was first trimmed off, then all images and calibration frames were run through a variant of the data reduction pipeline originally developed for processing
Wide Field Camera (WFC) data from the Isaac Newton Telescope (INT) -- for further
details see \cite{Irwin1985, Irwin1997}, \cite{Irwin2001} and \cite{Irwin2004}.

Prior to deep stacking, detector-level catalogues were generated for each 
individual processed science image to further refine the astrometric
calibration and also to assess the data quality. For astrometric calibration,
a Zenithal polynomial projection \citep{Greisen2002} was used to
define the World Coordinate System (WCS). A fifth-order polynomial includes 
all the significant telescope radial field distortions leaving a 
six-parameter linear model per detector to completely define the 
required astrometric transformations. The TwoMicron All Sky Survey (2MASS) 
point-source catalogue \citep{Cutri2003} was used for the astrometric reference 
system.

Quality control assessment for each exposure was based on the average seeing 
and ellipticity of stellar images, together with the sky level and sky noise, 
all determined from the object catalogues. During the stacking process the 
individual MegaCam catalogues were used, in addition to the standard WCS 
solution, to further refine to the sub-pixel level the alignment of the 
component images.  The common background regions in the overlap area from
each image in the stack were used to compensate for sky variations
during the exposure sequence and the final stacks for each band included seeing
weighting, confidence (i.e. variance) map weighting and clipping
of cosmic rays.

As a final image processing step, catalogues were derived from the deep
stacks for each detector and their WCS astrometry was updated. All objects 
detected in the catalogues are morphologically classified
(stellar, non-stellar, noise-like) before creating the final
band-merged $g$ and $i$ products.  The catalogues provide additional
quality control information and the classification step also computes
the aperture corrections required to place the photometry on an
absolute scale.  The band-merged catalogues for each field are then
combined to form an overall single entry $g$, $i$ catalogue for each
detected object. In this process, objects lying within 1 arcsec of each
other are taken to be the same and the entry with the higher
signal-to-noise measure is retained.  Objects present only on $g$ or
$i$ are retained throughout this process.

In a further refinement, we also used the better seeing $i$ band data to 
generate a list-driven version (i.e. forced photometry) of the $g$ band 
catalogue.  This catalogue was used to provide an alternative band merged product
for further analysis.

In the following, unless otherwise stated, the magnitudes are presented in 
their natural instrumental (AB) system with reddening corrections derived star
by star from the \cite{Schlegel1998} dust extinction maps combined with
the \cite{Schlafly2011} extinction coefficients for the $g, i$ bands.
We use the following correction coefficients: $g_o=g-3.793E(B-V)$ and 
$i_o=i-2.086E(B-V)$ (\citealt{Schlegel1998}).

\subsection{The Sagittarius dwarf irregular galaxy}

\begin{figure*} %remove stars for half page
%\vspace{1pt}
\begin{center}
\includegraphics[width=1.0\linewidth]{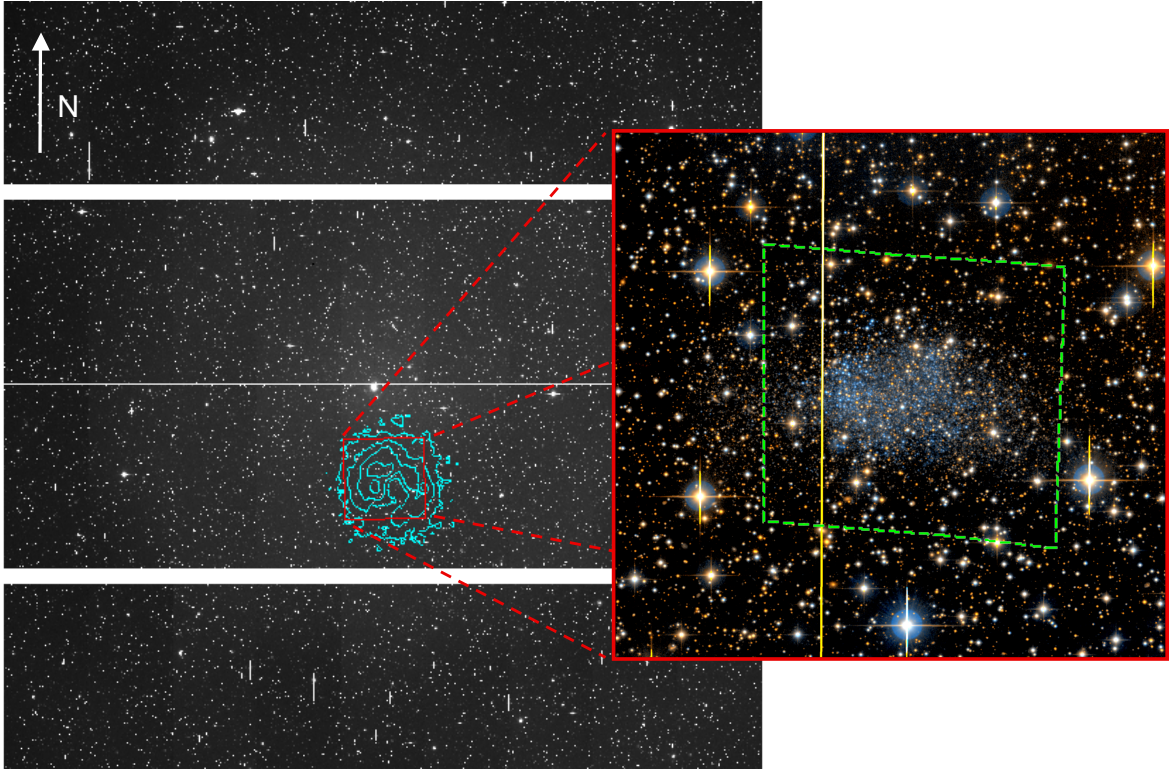}
\caption{Processed i band CFHT/MegaCam image, showing the full 1$ ^{\circ}$ by 1$ ^{\circ}$ field of view, oriented with north upward, east left. Overlaid blue contours show the HI distribution from LITTLE THINGS (Hunter et al. 2012), with contours at 10, 50, 100, 250, 500 (Jy/beam)(m/s). The inner 6' by 6' region is shown in the composite $g$ and $i$ band image,  with the green dashed box indicating  the HST ACS observations analyzed by Momany et al. (2005).} 
\label{FOV} 
\end{center}
\end{figure*}

We now turn our attention to the first galaxy analyzed as part of this program. Sag DIG is located near the zero velocity surface of the Local Group, where the gravitational attraction of the Local Group is balanced by the local Hubble flow. \cite{McConnachie2012} estimated that its free-fall timescale to either the MW or M31 is on the order of the age of the Universe, hence interactions with these systems seem improbable. The closest known galaxy to SagDIG is the low mass Aquarius dwarf galaxy (DDO210), still some 300 -- 400\,kpc away from SagDIG. 
Figure \ref{FOV} shows the processed $i$ band MegaCam image and an enlarged $g,i$ composite image of the central 6' by 6' region. Also overlaid in Figure ~\ref{FOV} are HI gas contours from the LITTLE THINGS survey (\citealt{Hunter2012}) and the HST ACS field of view studied by \cite{Momany2005} which is discussed later.

Sag DIG is classified as a dwarf irregular system. Figure~\ref{FOV} clearly shows blue stars visible in the center of the object, with fainter, redder stars more widely distributed. Numerous, often saturated, foreground Milky Way stars are obvious across the entire field of view, reflecting the fact that Sag DIG is located at moderately low Galactic latitude ($b = -16.3^\circ$). 

\cite{Cesarsky1977} first discovered Sag DIG on photographic plates obtained with the  ESO Schmidt telescope, and then studied it further with the ESO 3.6m and the Nan\c{c}ay radio telescope. Young blue stars were observed, as well as 21cm detections of neutral hydrogen, leading to this galaxy's classification as a dwarf irregular. The optical body of Sag DIG was noted to have a roughly elliptical shape, with approximate dimensions of  2.5' by 5'. \cite{Cesarsky1977} estimated its distance to be similar to that of NGC 6822 due to their close angular separation on the sky and similar  heliocentric velocities (-58 $\pm$ 5 km s$^{-1}$ for Sag DIG as compared to 57  km s$^{-1}$ for NGC 6822). A more precise estimate of its distance modulus was made using a comparison of its brightest blue stars to those in Sculptor and Phoenix, resulting in an estimate of $(m-M)_{AB}=25 \pm 1$. The mass to light ratio of $M(HI)/L_B=4.5 M_{\odot}/L_{\odot}$ was notably high, particularly since Sag DIG was one of the smallest and faintest galaxies at the time of its discovery. Subsequent observations by \cite{Longmore1978}  observed the neutral hydrogen with the Parkes 64m Telescope and broadly confirmed the findings of \cite{Cesarsky1977}. No resolved HII regions were detected, and the high  $M(HI)/L_B$ ratio was confirmed. 

Although \cite{Longmore1978} was originally unable to detect any HII,  two compact and one extended HII sources were observed via long slit spectroscopy using an H$\alpha$ map from \cite{Skillman1989}. The compact sources were Galactic in origin while the extended source belongs to Sag DIG. The oxygen abundance was determined to be approximately 3\% of solar. This value in in general agreement with the luminosity -- abundance relationship for such a faint system. \cite{Strobel1991} also studied the HII regions in Sag DIG and other dwarf irregulars, and found that across their sample a similar morphology  for all the HII regions.
 More recently, \cite{Saviane2002} studied O, N, Ne abundances in the brightest HII region of Sag DIG, finding $12+log(O/H)=7.26 - 7.50$, establishing Sag DIG's very low O abundance. 

A more recent study of the HI content of Sag DIG was conducted  by \cite{Lo1993} using VLA observations to study in detail a total of 9 dwarf irregulars including Sag DIG.  The HI was noted as being more extended than the stars and having a large crescent shape, in contrast to the very regular, elliptical, optical image of Sag DIG. In all the dwarfs studied by \cite{Lo1993} including Sag DIG, the HI is clumpy and not well described as a thin, disk-like structure as is often found in more luminous galaxies. Pressure support is generally more important than rotational support for these low mass systems. These observations are supported by \cite{Young1997}, where VLA and single dish observations of Sag DIG show no systematic rotation, making it hard to estimate the mass of the galaxy. The HI gas is found to have two components, one with a broad velocity dispersion of 10 km s$^{-1}$ distributed throughout the galaxy and a narrow component with a velocity dispersion of 5 km s$^{-1}$ found in clumps with masses on the order of $8\times 10^5 M_{\odot}$. \cite{Young1997} also compare the stellar and HI features in detail. They estimate that the HI is about three times more extended than the stellar component, and that the HI clumps do not correlate with any features in the stellar distribution. The position of one of the HI clumps correlates well with the previously identified HII region, suggesting it is potentially a star forming region. \cite{Young1997} also compute timescales for the collapse of the HI gas, considering that it is not rotationally supported, and suggest that magnetized turbulence is a plausible explanation to help support the gas and prevent more rapid collapse. Much of the structure of the gas discussed by these authors is visible in the LITTLE THINGS HI map (\citealt{Hunter2012}) shown in Figure~\ref{FOV}.

\cite{Karachentsev1999} obtained resolved stellar photometry in the V and I bands for Sag DIG. An updated distance modulus was found using the tip of the RGB to be $(m-M)_o =25.13\pm 0.2$, corresponding to $1.06\pm 0.10$ Mpc  and placing Sag DIG on the edge of the Local Group. Sag DIG's low metallicity was confirmed by \cite{Karachentsev1999} using the calibration of the RGB color as done for globular clusters. They found a mean metallicity is [Fe/H]=$-2.45\pm0.25$\,dex. This is more metal poor than the estimate of 3\% of Solar from \cite{Skillman1989} based on estimates from HII regions. However, given these two tracers probe very different ages in the galaxy, these estimates are not in obvious tension. \cite{Karachentsev1999} found an exponential scale length of 27."1, a central surface brightness of $\mu_V=23.9$ mag. arcsec$^{-2}$ and an absolute magnitude of $M_V=-11.74$. The star formation history was estimated by generating a synthetic CMD, showing that Sag DIG's current star formation rate is high relative to its mean. Overall, \cite{Karachentsev1999} noted that Sag DIG appears to be a ``normal" gas rich, low metallicity dwarf irregular.  \cite{Lee2000} conducted a similar analysis to that of \cite{Karachentsev1999} using BVRI photometry from the 2.2m University of Hawaii Telescope on Mauna Kea and broadly confirmed the results of \cite{Karachentsev1999}, although they derived a larger exponential scale radius of $r_e=37"\pm2"$. They also noted that the youngest stars in their CMD -- some consistent with forming as recently as 10Myrs ago -- are very much more centrally concentrated than the other stellar populations.

As well as the presence of very young stars, an extended star formation history of Sag DIG was implied by the presence of carbon stars, first photometrically identified by \cite{Cook1987}. Two populations of carbon stars were identified, one of which is brighter and redder, the other is fainter and bluer. The redder population are younger, higher metallicity carbon stars like those observed in the Large and Small Magellanic Clouds, while the bluer population is more like carbon stars in the dwarfs spheroidals, older and with a lower metallicity \citep{Cook1987}. The presence of both these populations suggest that Sag DIG has had multiple episodes of star formation (or prolonged star formation) at intermediate ages. 

\cite{Momany2002} studied the stellar populations of Sag DIG using deep BVI observations, and identified an AGB population as well as the RGB and blue loop stars, consistent with the observations of Carbon stars. A follow-up study of Sag DIG using the Hubble Space Telescope (HST) by  \cite{Momany2005} presented a much deeper CMD, albeit in a more restrictive field of view (see Figure~\ref{FOV}). They estimated a metallicity  for the galaxy between [Fe/H]=-2.2 to -1.9 dex. Within the CMD, they were able to identify a very old population (\textgreater 10 Gyrs) as well as a red clump indicative of intermediate ages. They also compared the locations of the stars to the HI gas, and found that the youngest stars are found near to, but not coincident with, the HI clumps.   \cite{Weisz2014} determined a star formation history (SFH) for Sag DIG by generating synthetic CMDs for a given stellar population and matching them to features observed in the HST CMD obtained by \cite{Momany2005}. However, the data does not reach as deep as the oldest main sequence turn-offs, and so there are very large uncertainties as to the fraction of stars formed at the oldest times. Nevertheless, it is clear that Sag DIG has had on going star formation continuing up until present, with about half its stellar mass formed by $\sim6$Gyrs ago. 

The most recent study of Sag DIG  -- published during the preparation of this manuscript -- is by \cite{Beccari2014}. They, like us, examined the extended structure of Sag DIG from wide field photometry by using resolved stars to trace the faintest features, and find that it is well described by a single exponential curve with a scale radius of order 340 pc, with no evidence of disturbances or breaks in the profile. We will return to discussion of this paper and compare it to our findings for Sag DIG in Section 5.
   
\section{Stellar Population Analysis}

\begin{figure*} %remove stars for half page
%  \vspace{100pt}
\begin{center}
\includegraphics[width=\linewidth]{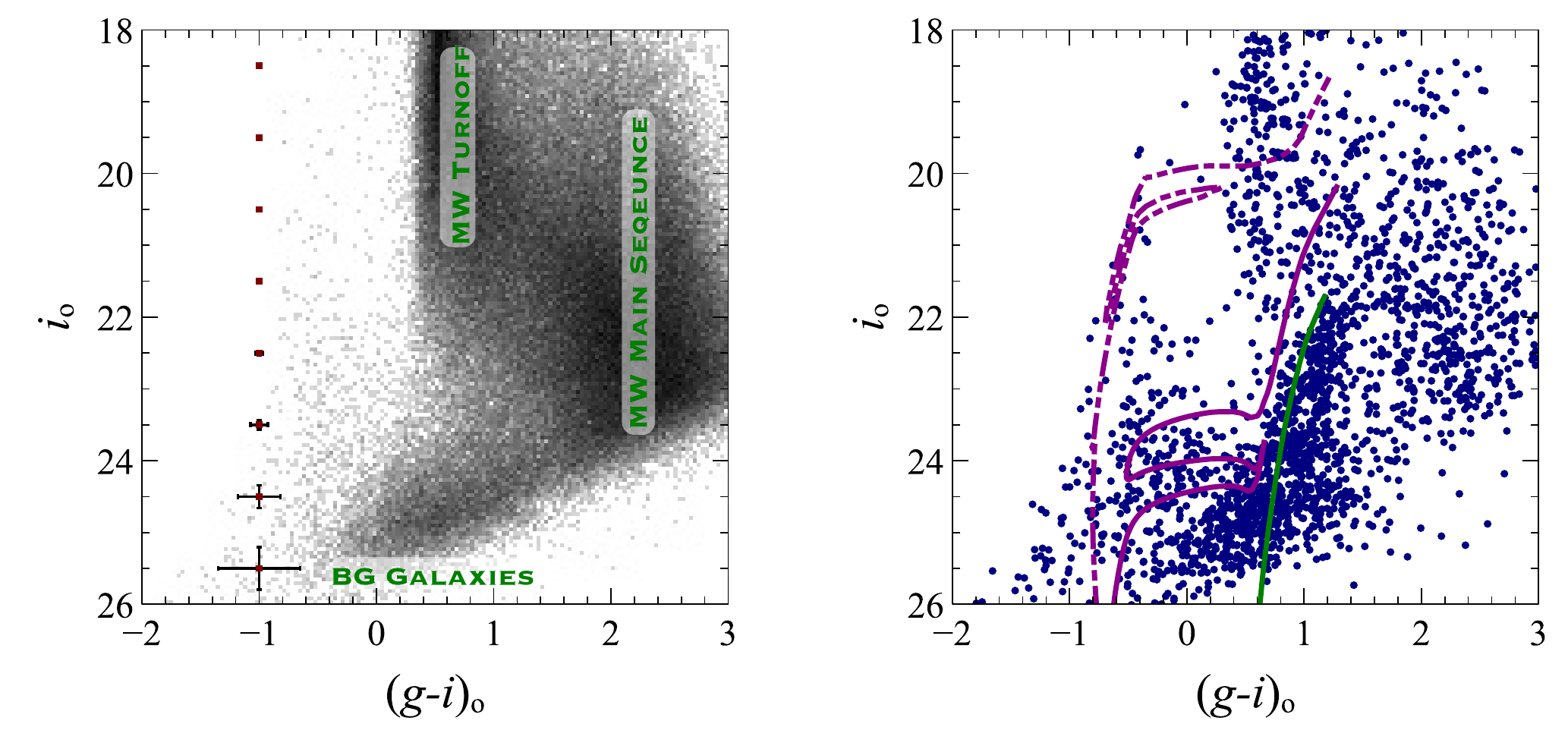}
\caption{\textit{Left Panel:} A Hess diagram (logarithmically scaled) of the CMD  for all stars in the CFHT/MegaCam field. Mean errors as a function of magnitude are shown. Regions of significant foreground and background contamination are labelled with the source of contamination. \textit{Right Panel:}  Blue points show the CMD of all stars within 4 $r_{s}$ from the center of Sag DIG.  The green isochrone traces a RGB, with an age of 6 Gyrs and [Fe/H]=-2.2 dex.  The purple isochrones are for younger populations with ages 50 Myrs (dashed) and 500 Myrs (solid) and a metallicity of [M/H]=-2.0 dex.  All isochrones are from the PARSEC isochrone set \citep{Bressan2012} and have been adjusted to the distance of Sag DIG as determined in Section \ref{distest}. }
\label{cmdfield}
\end{center}
\end{figure*}

The outskirts of Sag DIG are resolved into stars in our CFHT/MegaCam imaging. We are able to construct a color-magnitude diagram (CMD) of these sources to identify the major stellar populations, estimate metallicity, and examine the influence of background and foreground contamination, among other applications. The left hand panel of Figure \ref{cmdfield} shows a Hess diagram of the Sag DIG CMD, using the full 1$^{\circ}$ by 1$^{\circ}$ field of view. The \cite{Schlafly2011} update of the \cite{Schlegel1998} dust maps are interpolated for each identified star to correct for foreground dust; the average correction in the $(g,i)$ bands is (0.45, 0.25)\,mags. The average error in both $i_0$ and $(g-i)_0$ is shown. Below a magnitude of $i_0\simeq25$, the uncertainty in the photometry becomes very large.

Sag DIG is at relatively low Galactic declination, and so the field is polluted with a large number of foreground Milky Way stars. Stars near the main sequence turnoff in the Milky Way halo, located at a range of distances, produce the vertical sequence with a color near $(g-i)_0$= 0.5,  labelled in the left panel of Figure \ref{cmdfield}.  Low mass stars in the Milky Way disk at varying distances appear in the cloud of objects around $(g-i)_o$= 2, and are also labelled.  An additional source of contamination of the CMD are distant (elliptical) galaxies misidentified as stars at faint magnitudes, and these dominate the CMD at the faintest magnitudes. To remove the overwhelming majority of this contamination, the CMD is restricted to stars within 4 $r_s$  of the center of Sag DIG (where $r_s$ is the characteristic radius from the Sersic fit to the radial profile, described in detail in Section \ref{fit}). The CMD is dominated by stars native to Sag DIG, shown in  the right hand panel of Figure \ref{cmdfield}

As introduced in the previous section, the SFH derived by \cite{Weisz2014} shows an extended period of star formation and appears that Sag DIG has a fairly ``typical" SFH as compared to other field dwarf irregulars. These galaxies generally form 30\% of their mass prior to z $\sim$2 and show an increase in star formation after z $\sim$ 1. Qualitatively, we can see various features suggesting an extended period of star formation; the prominent RGB suggests an old (at least \textgreater 2 Gyrs) population while the bright blue stars indicate recent star formation on the timescale of \textless 1 Gyr.   PARSEC \citep{Bressan2012} isochrones  are shown in Figure~\ref{cmdfield} shifted to the distance of Sag DIG. 
The isochrone overlaid on the RGB  is  6 Gyrs -- corresponding to the mean age of Sag DIG based on the analysis by  \cite{Weisz2014} -- and metal poor, [Fe/H]=-2.2 dex. The fact that we see a significant number of RGB stars lying blue-ward of this isochrone are indicative of either an even more metal poor population or a younger component to the RGB; the degeneracy between age and metallicity makes interpreting the colour distribution of RGB stars uncertain (we return to this point in Section \ref{metest}). While we account for reddening due to Galactic dust, no correction for internal extinction is included; however, this effect would make stars appear redder and so would not explain the position of the RGB stars to the blue side of the isochrone in Figure \ref{cmdfield}.
Clearly, Sag DIG has a metal poor component and has had multiple, extended periods of star formation, consistent with the previous results discussed in Section 2.

We refer the reader to Weisz et al. (2014) for a more detailed interpretation of the stellar populations of Sag DIG based on deep HST imaging. While the wide-field aspect of the {\it Solo} data enables a global analysis of the stellar populations of this and other systems, the depth of the photometry is relatively shallow in comparison to what has been achieved for individual Local Group galaxies using HST (especially, the comprehensive study by \cite{Weisz2014} and the LCID project; \citealt{Gallart2015}). Thus with {\it Solo}, we do not we do not seek to perform this type of analysis on an individual basis, but instead defer to a study of the entire population of {\it Solo} systems to be presented in a future paper, with the focus on the statistical properties of the stellar populations/star formation histories of the outer regions. 

\subsection{Distance Estimate} \label{distest}

\begin{figure} %remove stars for half page
%  \vspace{100pt}
\begin{center}
\includegraphics[width=\linewidth]{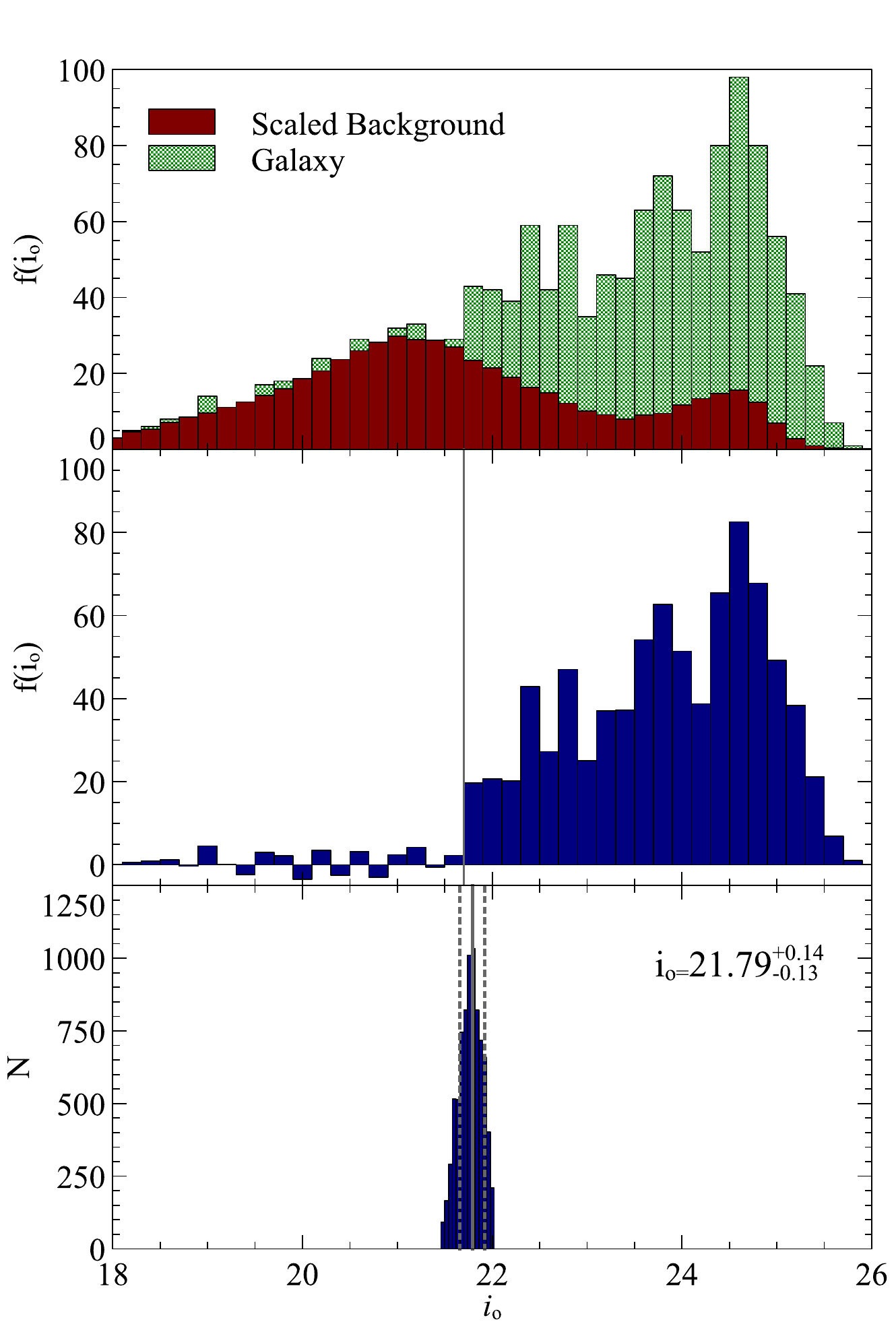}
\caption{The upper panel shows the luminosity function for both Sag DIG (selected to be within $6r_s$ from the galaxy center) and the field (selected to be $>10r_e$ from the galaxy center), scaled by area. The middle panel shows the background subtracted luminosity function.  The distribution of values calculated for the TRGB is shown in the lower panel, derived using the method described in the text.  }
\label{3plot}
\end{center}
\end{figure}

We use the tip of the red giant branch (TRGB) as a standard candle to determine the distance to Sag DIG. The luminosity of RGB stars as they undergo the helium flash and begin fusing helium in their cores is approximately constant, largely independent of stellar age and metallicity, at least for ages greater than $\sim 4$ Gyrs and a metallicity lower than [Fe/H]$\simeq-1.0$ dex (see \cite{Lee1993}, \cite{Bellazzini2001},  \cite{Madore2009} and \cite{Salaris1997} amoung others). Assuming that the RGB is well populated, the maximum luminosity of the RGB stars is characteristic of the helium flash and can be identified by looking for a large discontinuity in the luminosity function (\citealt{Sakai1996}). AGB stars, if present, can slightly mask this discontinuity, although the scale of the effect for Sag DIG will be tiny due to the relatively weak AGB population and the very strong RGB population. 

The luminosity function for the RGB stars in Sag DIG is shown in the top panel of Figure \ref{3plot}, constructed from stars bounded by a set of ``tramlines'' that were drawn around the RGB in the right panel of Figure \ref{cmdfield}. The luminosity function is corrected for foreground/background contamination by constructing a similar luminosity function for stars within the same region of the CMD but that are located at least 10 $r_{s}$ from the center of Sag DIG (where the foreground/background dominates). The resulting luminosity function is scaled by area and also shown in the top panel of Figure~\ref{3plot}. The corrected, subtracted luminosity function is shown in the middle panel of Figure~\ref{3plot}. 

The TRGB is clearly visible in the middle panel of Figure~\ref{3plot} as the large jump in star counts at $i_0 \sim 21.8$. To more precisely determine the TRGB magnitude and uncertainties, we follow \cite{Sakai1996} and use a 5 point Sobel filter on the luminosity function in Figure \ref{3plot}. We repeated our analysis with a three point Sobel filter with no change in the result. To remove the dependency on choice of bin edges and bin sizes, we conduct this analysis on multiple realizations of the luminosity function: the bright magnitude end was varied from $i_0=20$ to $i_0= 21$ in 2000 steps, and for each of these realizations the bin width was varied from 0.17 to 0.25 mags in 5 steps. This combination resulted in 10 000 luminosity functions on which the Sobel filter was applied to find the TRGB. The distribution of results is shown in the lower panel of Figure \ref{3plot}. The mean value from this distribution and  bounds containing 68\% of the results are used as the luminosity of the TRGB and associated uncertainties, such that $i_{TRGB} = 21.79^{+0.14}_{-0.13}$.

Using the PARSEC isochrones \citep{Bressan2012} in the CFHT filter system, the luminosity of the TRGB is $i_0$=3.53 for an old (approximately 12 Gyrs) and metal poor population. The resulting distance modulus of Sag DIG is therefore $(m-M)_0$=25.36$_{-0.14}^{+0.15}$, corresponding to a distance of $1.18^{+0.08}_{-0.06} $ Mpc. This result is in agreement with recent estimates by \cite{Momany2002} ($(m-M)_0=25.14 \pm 0.18 $) and \cite{Momany2005} ($(m-M)_0=25.10 \pm0.11$), which implement different reddening corrections, also including internal extinction. 
It is also in agreement with the most recent estimate by \cite{Beccari2014}, finding $(m-M)_0$ = 25.56$\pm0.11$.

\begin{figure} %remove stars for half page
\begin{center}
\includegraphics[width=\linewidth]{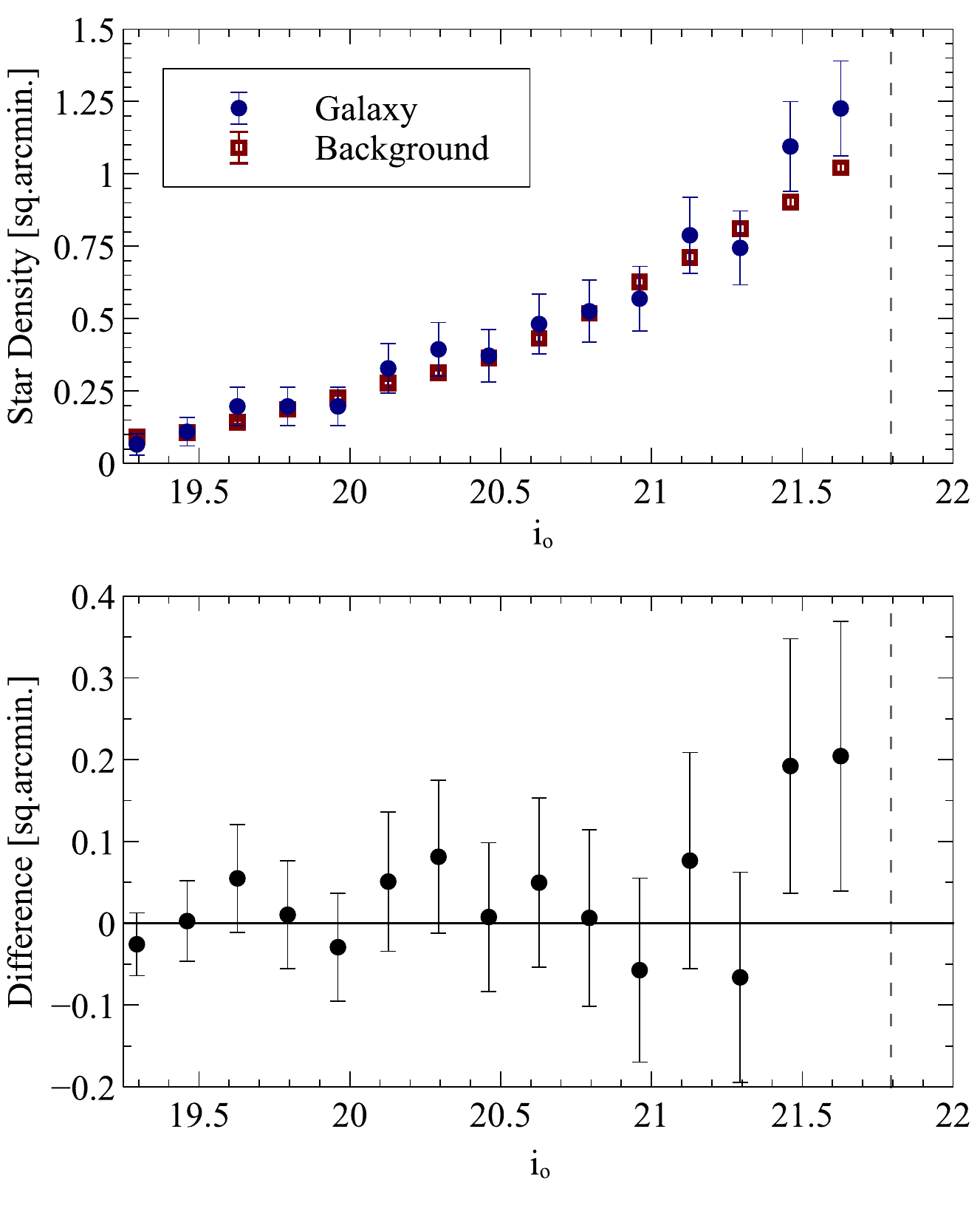}
\caption{The upper panel shows the luminosity function of stars above the TRGB within the tramlines used earlier, for objects within the Sag DIG field and the (scaled) reference field (blue and red, respectively), in $\Delta i_o = 0.125$ mag bins. The lower panel shows the corrected luminosity function, indicating an excess of stars in the Sag DIG field that we attribute to an AGB population with the brightest star at  $i_0\simeq20.25$ . The dashed line indicates the location of the TRGB.}
\label{agbfig}
\end{center}
\end{figure}

We reexamine the CMD and luminosity function more closely for evidence of an AGB population. Both the carbon stars identified by Cook (1987) and the later analysis by Momany et al. (2002) establish that Sag DIG has a AGB population, although a bright AGB population above the TRGB is not clearly visible in the CMD of Figure 3 or in the luminosity function in Figure 4. As such, we zoom in to the region of the luminosity function immediately brighter than the TRGB by counting the number of stars in  $\Delta i_o = 0.17$ dex bins above the TRGB within the tramlines that were used in Section 3.1, in both the CMD of Sag DIG and the reference CMD. The number of stars in these bins as a function of magnitude is shown in the top panel of Figure 5, along with their Poisson uncertainty. The subtracted counts are shown in the lower panel, again with the uncertainties. We can see that  there is essentially no excess of stars in the Sag DIG CMD brighter than the TRGB relative to the reference field. The two faintest bins both show an excess, potentially indicative of an AGB population but, as the error bars demonstrate, it is a weak population if it is in fact real. Given the low number of bright AGB stars that we can detect in our luminosity function, particularly in comparison to the RGB, the TRGB remains a reliable distance indicator.

\subsection{Metallicity Estimates}\label{metest}

\begin{figure} %remove stars for half page
%  \vspace{100pt}
\begin{center}
\includegraphics[width=\linewidth]{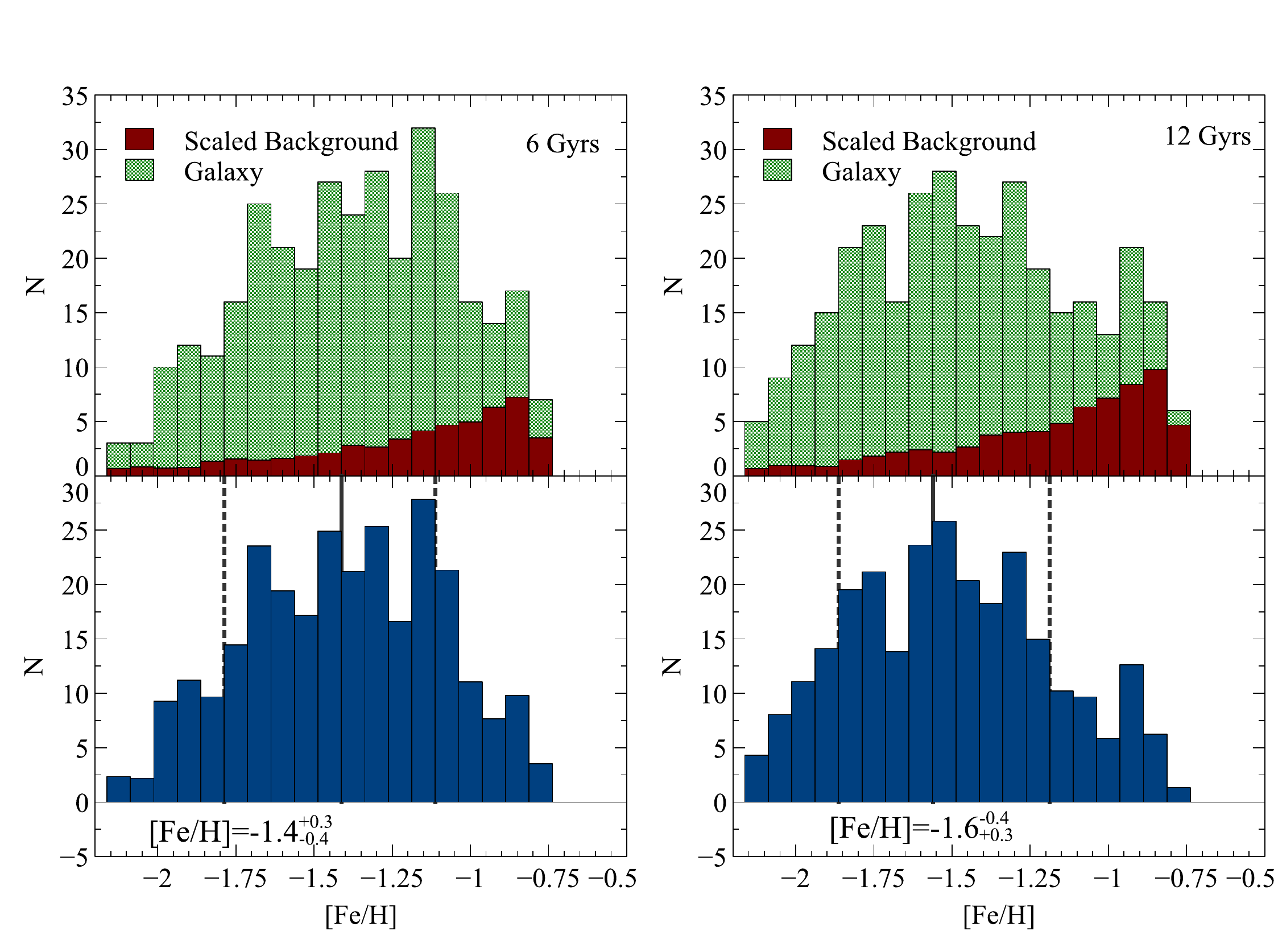}
\caption{The MDF for the galaxy based on stars within $4r_s$ from the center of Sag DIG is shown in the top panels in green. The scaled MDF for the background field, consisting mostly of foreground Milky Way stars, is shown in red. The subtracted MDF is shown in the bottom panels. On the left, the analysis is conducted using 6 Gyrs isochrones, whereas on the right, a 12 Gyrs set of isochrones is used. The solid line indicates the median value, while the dashed lines represent interquartile range between 16\% and 84\%, approximating one sigma.  We lack of knowledge of the distribution more metal poor than [Fe/H] $\sim2.2$, since there are a significant numbers of stars with colors bluer than the bluest isochrone ([Fe/H]=-2.2 dex).}
\label{mdf}
\end{center}
\end{figure}

The metallicity distribution function (MDF) for the red giant branch of Sag DIG, shown in Figure \ref{mdf},  is determined using a bilinear interpolation of stellar position in the CMD between a set of PARSEC isochrones with a fixed age (e.g. \citealt{Bellazzini2003}, \citealt{Sarajedini2005}, or \citealt{Ordonez2015}).  A set of 12 Gyrs isochrones with metallicities ranging from [Fe/H]=-2.2 to -0.3 dex were used as well as a 6 Gyr set (corresponding to the approximate mean age of Sag DIG derived by \cite{Weisz2014} as discussed previously). 
For simplicity, all isochrones used have [$\alpha$/Fe]=0.0 dex. To minimize uncertainties in the MDF caused by stellar photometric errors, and also to reduce the contamination due to misidentified galaxies (that becomes more common at fainter magnitudes), only stars brighter than $i_o$=24 are used. In the same way as for the luminosity function in the previous section, a Sag DIG MDF and a reference field MDF is created are by selecting stars within $4r_{s}$, and further away than $10r_{s}$, respectively. The solid line in Figure \ref{mdf} indicates the median metallicity for each set of isochrones, while the dashed lines denote a percentile range from 16\% to 84\%, approximating one sigma deviation from the median value. For the 6 Gyr isochrones, the median metallicity is [Fe/H]=-1.4$^{+0.3}_{-0.4}$ dex while for the 12 Gyr set the median metallicity is  [Fe/H]=-1.6$^{+0.4}_{-0.3}$ dex. For comparison, the same analysis is performed using the Dartmouth isochrone set from \cite{Dotter2008}, which are known to describe the RGB branch well in the CFHT/MegaCam filter system \citep{McConnachie2010}. The resulting MDF is very similar, with the resulting median metallicity of [Fe/H]=-1.4$\pm0.5$  dex and [Fe/H]=-1.6$^{+0.7}_{-0.4}$ dex for 6 Gyrs and 12 Gyrs respectively. 

As previously discussed, Sag DIG is not well described by a stellar population with a single age, and so the basic assumption that goes into the derivation of the MDFs is flawed. However, the adoption of a few different age assumptions helps to estimate the systematic uncertainty produced by this assumption, and it appears to be of order a few tenths of a dex. Also of particular note, there are a significant number of stars that appear bluer than the most metal-poor isochrone ([Fe/H]$ = -2.2$\,dex) for both age assumptions. Due to the way in which we conduct the bilinear interpolation to estimate metallicities, these blue stars are not included in the analysis, thus we are clearly overestimating the metallicity for Sag DIG.  Hence, the metallicity estimate is, at best, an upper limit and explains why the derived mean metallicity is of order [Fe/H]$ \sim -1.5$\,dex, considerably more metal rich than previous estimates by \cite{Karachentsev1999} and \cite{Lee2000}. From comparing the mean color of the RGB to the isochrones in the PARSEC isochrone set, it appears a more realistic metallicity estimate for Sag DIG is of order [Fe/H] $\simeq - 2.2 $ dex (for example, see the isochrone in Figure~\ref{cmdfield}).

The presence of so many very blue RGB stars in Sag DIG is curious. In particular, while it is possible that their color is a result of a very low metallicity, the color-metallicity relation for RGB stars at these metal-poor values quickly ``saturates", and more metal poor stars are not notably much bluer under normal circumstances. Therefore, the RGB possibly contains giants that are not well described by intermediate or old  RGB isochrones, for example either young giants or AGB stars.  Both types of giants are possible for a galaxy that has ongoing star formation (supported by \citealt{Weisz2014}), and may suggest that the metallicity is therefore not as low, on average, as the RGB color would seem to suggest. Clearly, spectroscopic estimates of stellar metallicities in this galaxy would be particularly interesting to obtain (for example, see the work on WLM by \citealt{Leaman2012}).

\subsection{Stellar Populations Gradients}

\begin{figure*}%remove stars for half page
%  \vspace{100pt}
\begin{center}
\includegraphics[width=0.75\linewidth]{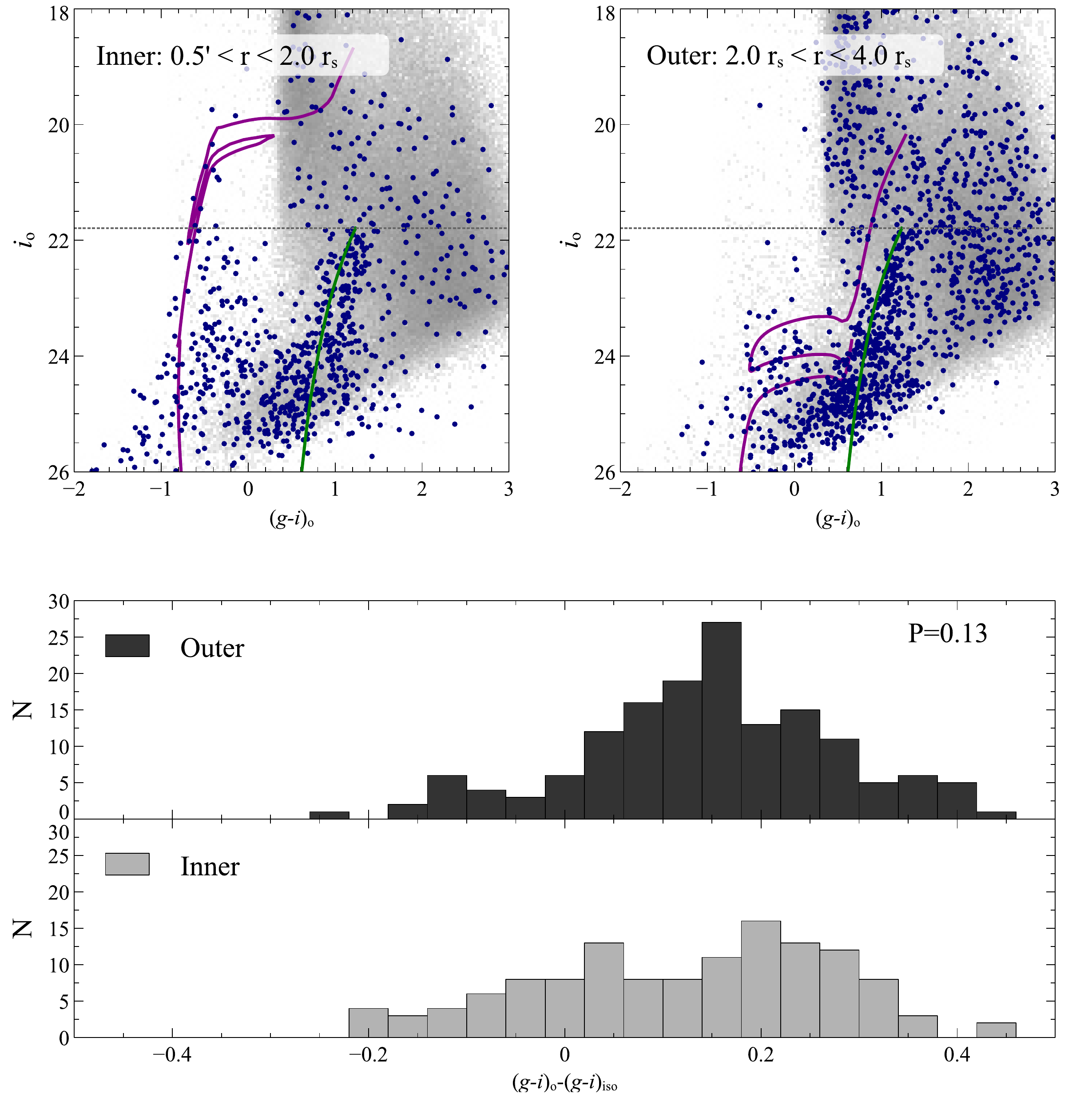}
\caption{The upper panels show the CMD for the inner part of the galaxy (defined to be within $2r_s$) on the left and  the outer part of  galaxy (defined to be between 2 to 4$r_s$) on the right shown relative to the 6 Gyr isochrone in green with a metallicity of [Fe/H]=-2.2 dex (\protect\citealt{Bressan2012}).  Younger isochrones are shown in purple with ages of 50 Myrs and 500 Myrs in the left and right panels respectively.   The color difference of RGB stars relative to the 6 Gyr isochrone is shown in the lower panels, along with the Kolmogorov-Smirnov test result between  the two color distributions.}
\label{mdfinout}
\end{center}
\end{figure*}

Thus far, our analyses assume that there is no variation in the stellar populations of Sag DIG over its spatial extent. However, a large body of work has demonstrated that there can be significant variation in the stellar populations of dwarf galaxies over their spatial extent (e.g. \citealt{Harbeck2001}, \citealt{Tolstoy2004}, \citealt{Rizzi2004}, \citealt{Battaglia2006}, \citealt{Lianou2013}, \citealt{Monelli2012}, \citealt{McMonigal2014}, \citealt{Bate2015}). We examine the populations of Sag DIG in this regard by splitting the galaxy into inner and outer sections. ``Inner'' is defined to be within $2r_s$ (excluding the inner 0.5' which is highly impacted by crowding) and ``outer'' is defined to be within 2 and 4$r_s$. The CMDs are shown in the left and right upper panels of Figure \ref{mdfinout}, respectively.

Inspection of the CMDs in Figure~7 shows that there is a much higher concentration of young stars (blue loop and main sequence) in the central regions of Sag DIG than the outer regions (the outer CMD spans an area 4 times larger than the inner CMD, yet still has fewer stars with $(g-i)_0\lesssim1$). Moreover, the outer region lacks very bright young blue stars in comparisons to the inner region, for example young stars brighter than $i\simeq 22$\,mags. Overlaying the younger isochrones shown previously in Figure \ref{mdfinout}, we can see that the brightest blue stars in the central regions correspond to a population approximately 50 Myrs old, whereas the brightest blue stars in the outer region are better fitted with an isochrone that is at least $\gtrsim100$\,Myrs old.

The RGB is well populated in both the inner and outer CMDs, however the color spread is larger in the inner region. In particular, there may be a deficit of blue RGB stars in the outer region (i.e. stars bluer than the reference isochrone 6 Gyrs, [Fe/H]=-2.2 dex). To quantify this difference and determine if it is significant, the color difference of each star from the reference isochrone is computed. Only stars between the TRGB and $i_o$=23.5 are included (to minimize uncertainties due to photometric errors), and only RGB stars are selected. These tramlines are chosen to be far enough from the RGB to generously include all possible RGB stars, so a small amount of foreground contamination will be present.
 
The lower panel of Figure \ref{mdfinout} shows that the color distribution of the RGB stars in the inner and outer regions are qualitatively different, with more stars at negative color difference (i.e., the blue side of the isochrone) in the inner region compared to the outer region. To quantify the potential difference between the populations, a Kolmogorov-Smirnov test is applied. The resulting $p$ value is reasonably large, hence we must conclude that any difference, if present, is not statistically significant. If the excess of bluer RGB stars in the center is real, then it implies either a younger RGB population or a more metal poor RGB population. A dwarf galaxy with a more metal poor interior would be quite unusual, since all radial gradients in metallicity so-far discovered act in the opposite direction (e.g., \citealt{Ross2015} or \citealt{Leaman2013}). We prefer the interpretation that the difference in color is due to much younger ages in the center, obviously consistent with the centrally concentrated star formation of this galaxy.

\section{Low Surface Brightness Structure}

\subsection{Integrated Light Profiles}\label{integrated}

While {\it Solo} targets galaxies for which some study of their resolved stellar content is possible, most of the galaxies have central regions that cannot be resolved from the ground under natural seeing conditions. Thus, analysis of the integrated light is also necessary for a complete view of the galaxy. Integrated light radial profiles are generated by finding the total flux within a set of elliptical annuli. For simplicity and ease of comparison, the ellipticity $e$ and position angle $PA$ of the annuli is fixed with radius. We adopt $e = 0.53$ and $PA = -0.52^{\circ}$ as determined later in Section \ref{2DDM}.  

\begin{figure} %remove stars for half page
\begin{center}
\includegraphics[width=\linewidth]{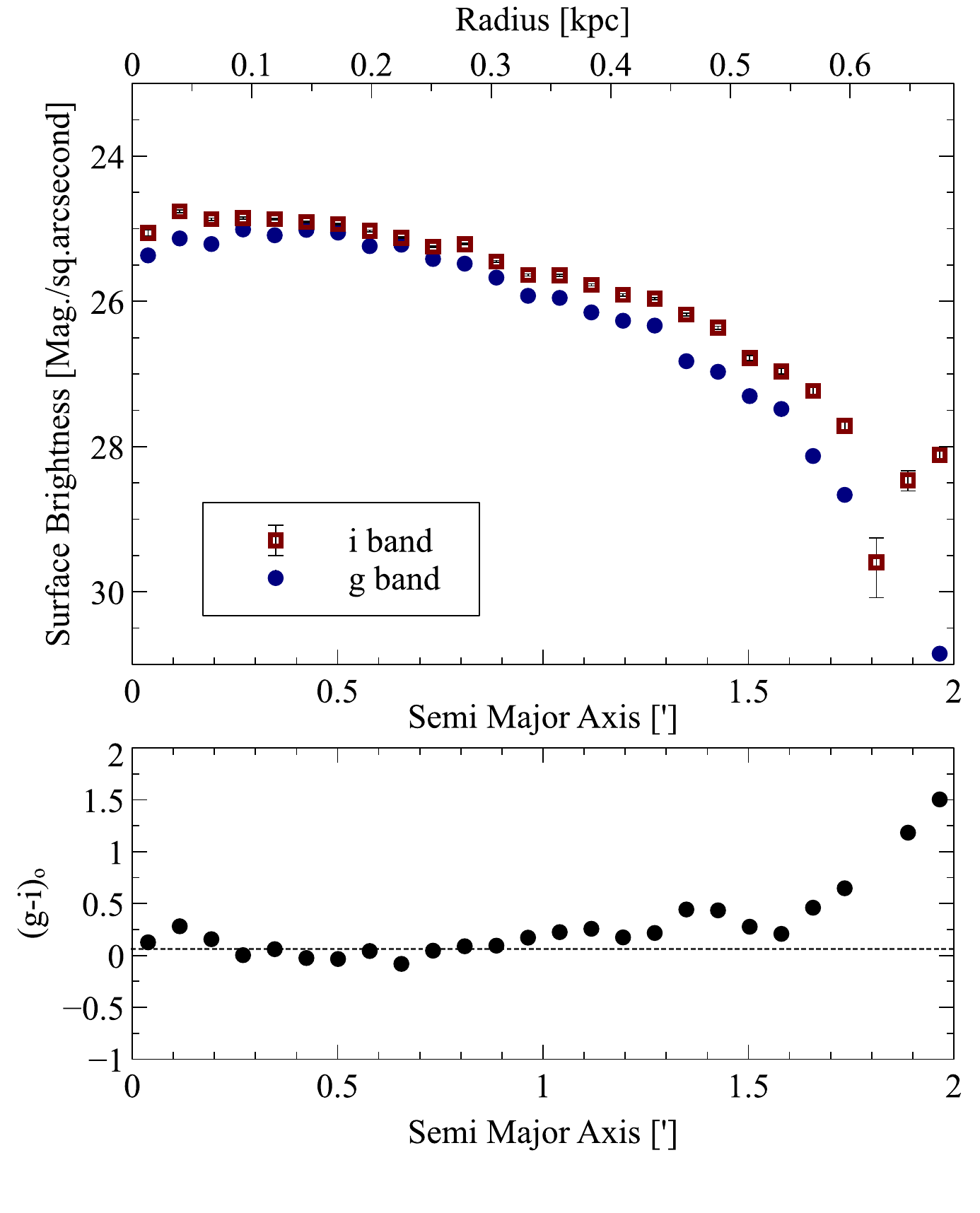}
\caption{The upper panel shows integrated light profiles in both the $i$ and $g$ bands, calculated using elliptical annuli with bright sources masked as described in the text. The lower panel shows the color profile as a function of radius, and the dashed line shows the mean color of the galaxy within 1 arcmin ($(g-i)_0 = 0.06$). }
\label{integratedlightprof}
\end{center}
\end{figure}

The foreground Milky Way stars contaminate the underlying light from Sag DIG and must be carefully subtracted as these foreground stars are significantly brighter than the diffuse signal from the dwarf galaxy. Before determining the integrated light profile, all pixels brighter than a cutoff level, or adjacent to one of these bright pixels, are masked. The appropriate cutoff level for each band is determined by analyzing the distribution of pixel values across the image and by studying the resulting mask, for which all obvious foreground stars should be blacked out. Then, the total flux within each elliptical annulus for unmasked pixels is summed, simply ignoring the masked pixels in both flux and area. The uncertainties associated with the flux are Poisson photon noise. The resulting integrated light profile is shown in the upper panel of Figure \ref{integratedlightprof} for both  bands, and the resulting $(g-i)_o$ profile is shown in the lower panel.  We have corrected the flux in each pixel for the appropriate value of extinction as derived from the \cite{Schlafly2011} dust maps at the position of that pixel.

In general, the profiles in both $g$ and $i$ are broadly exponential with no notable features.

There is a possible ``kink" in both profiles near 1 arcmin which is most likely due to the irregular distribution of young stars, visible in Figure \ref{FOV}.
 From the color profile, we can see a small gradual reddening towards larger radii, particularly beyond $\sim 1$ arcmin, consistent with the fact that bluer stellar populations are more centrally concentrated. The dashed line in the lower panel is the average color of the galaxy interior to 1 arcmin  ($(g-i)_0 = 0.06$).

\subsection{The outer low surface brightness structure} \label{2DDM}

Whereas the integrated light allows us to probe the central regions of Sag DIG to a moderately low surface brightness level of $\sim 28$\,mags/sq.arcsec, we can use resolved star counts as a proxy to characterize the low surface brightness shape and extent of Sag DIG. Specifically, we use RGB stars selected from the CMD to study the low surface brightness structure. A caveat on this approach, however, is that the RGB distribution  probes the structure of the galaxy based on the presence of intermediate and old stellar populations, whereas the integrated light used in the previous section is a luminosity-weighted mapping of the structure of the galaxy (and is generally biased to the younger, brighter populations present in Sag DIG).

Prior to analyzing the stellar distribution, a correction must be applied for the large gaps between the CCDs in the CFHT/MegaCam focal plane, since star counts are artificially low (zero) in these regions. We generate  artificial RGB stars within gaps by binning the spatial distribution of RGB stars into pixels with a size 0.6' $\times$ 0.6' and identifying those pixels with anomalously low star counts (more than $\sim 3 \sigma$ lower than the mean of a column). Artificial stars are added at random positions with these pixels until the total number of stars in this pixel is equal to the interpolated value based on its neighbours, allowing for Poissonian scatter. We use these artificial stars only when constructing 1-D and 2-D density distributions of the RGB maps.

\begin{figure*} %remove stars for half page
%  \vspace{100pt}
\begin{center}
\includegraphics[width=\linewidth]{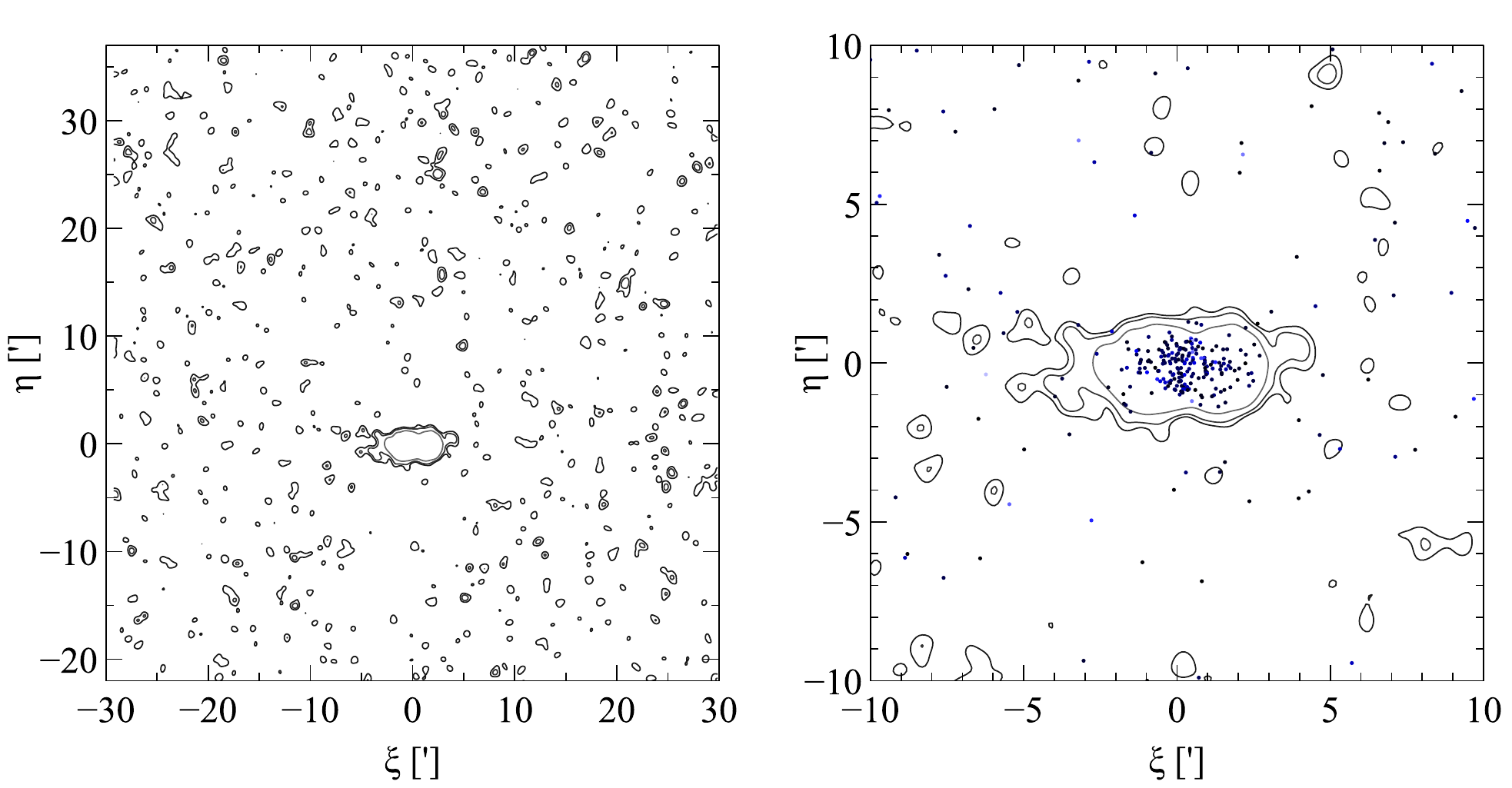}
\caption{The left panel shows the RGB star distribution (RGB stars per unit area, relative to the median background) in the full 1$^\circ$ by 1$^\circ$ image. The right panel shows in the inner region of the RGB stellar distribution with blue stars (selected with $-1<(g - i)_o < 0$ and $i_o$ \textless 25 mags.) plotted as points. Contours: 1.5, 3 and 10 $\sigma$ above the mean background. Bins are 0.1' and are smoothed using a gaussian filter with $\sigma=0.3$\,arcmins. }
\label{rgbstruc}
\end{center}
\end{figure*}

RGB stars are selected from the CMD using the same selection criteria as used when deriving the luminosity function. An image of the density distribution of these stars is created using pixels with dimensions $0.1' \times 0.1'$. A  Gaussian filter is applied, with $\sigma=0.3$\,arcmins. The mean background value and its standard deviation is estimated from the outskirts of the image (a border with width 10 arcmins), and that is used to set appropriate contour levels. The resulting RGB density map is shown in the left panel of Figure \ref{rgbstruc} for the entire CFHT/MegaCam field, with contour levels at $1.5, 3$ and $10\,\sigma$ above the background. The right panel shows an enlargement of a $20 \times 20$ arcmins region centered on Sag DIG. Blue points are the positions of young blue stars, $-1<(g - i)_o < 0$ and $i_o$ \textless 25 mags., for comparison.

\begin{figure*} %remove stars for half page
%  \vspace{100pt}
\begin{center}
\includegraphics[width=\linewidth]{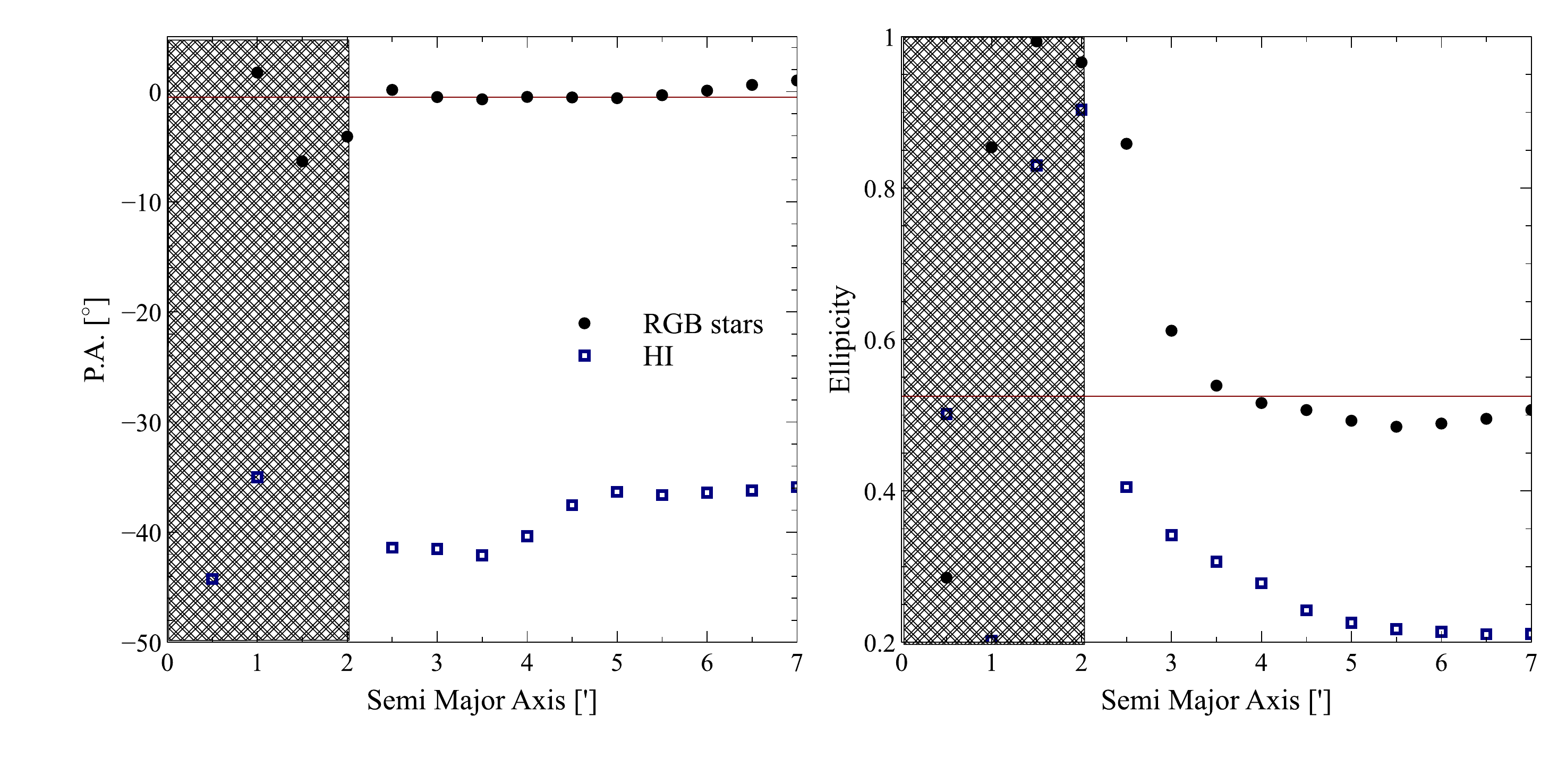}
\caption{ \textit{Left:}  The position angle as a function of the semi major axis, as described in the text. \textit{Right:}  The ellipticity as a function of the semi major axis, as described in the text. The hatched regions in both indicate the region where the estimates are heavily influenced by crowding. The lines indicate the mean values for the shape parameters of Sag DIG adopted for this study.}
\label{paell}
\end{center}
\end{figure*}

Figure~\ref{rgbstruc} shows that the outer regions of Sag DIG is well described as a highly elliptical system. The faint extension are at a level of approximately 1.5 -- 3$\sigma$ above the background level. We estimate the value for the ellipticity and position angle as a function of radius using the moments of the stellar distribution. As crowding is a significant issue in the central part of Sag DIG, we fix the center of Sag DIG at the literature value of $\alpha=19^{h}29^{m}59.0^{s}$ and $\delta = -17^{d}40^{m}41^{s}$ \citep{McConnachie2012}. At a fixed distance from Sag DIG, we use all pixels that are within that distance to calculate the moments of the distribution following \cite{McConnachieIrwin2006} and hence the position angle and ellipticity. We then repeat the analysis, but this time using only those pixels contained within an ellipse with the newly derived ellipticity and position angle. This process is repeated until convergence. The resulting position angle and ellipticity are shown as a function of major axis radius in Figure \ref{paell}. The hatched regions of these plots correspond to the approximate regions where crowding is problematic and so the estimates are unreliable. Solid lines indicate the adopted mean position angle and ellipticity of the outer points, between a semi-major axis of  2.5' and 7.5'.  We find that $e=0.53\pm 0.04$ and $PA \simeq -0.52^{\circ}\pm0.14^{\circ}$ with very little radial variation away from the center of the dwarf.

\subsection{Radial profiles to low surface brightness limits}\label{fit}

\begin{figure*} %remove stars for half page
%  \vspace{100pt}
\begin{center}
\includegraphics[width=\linewidth]{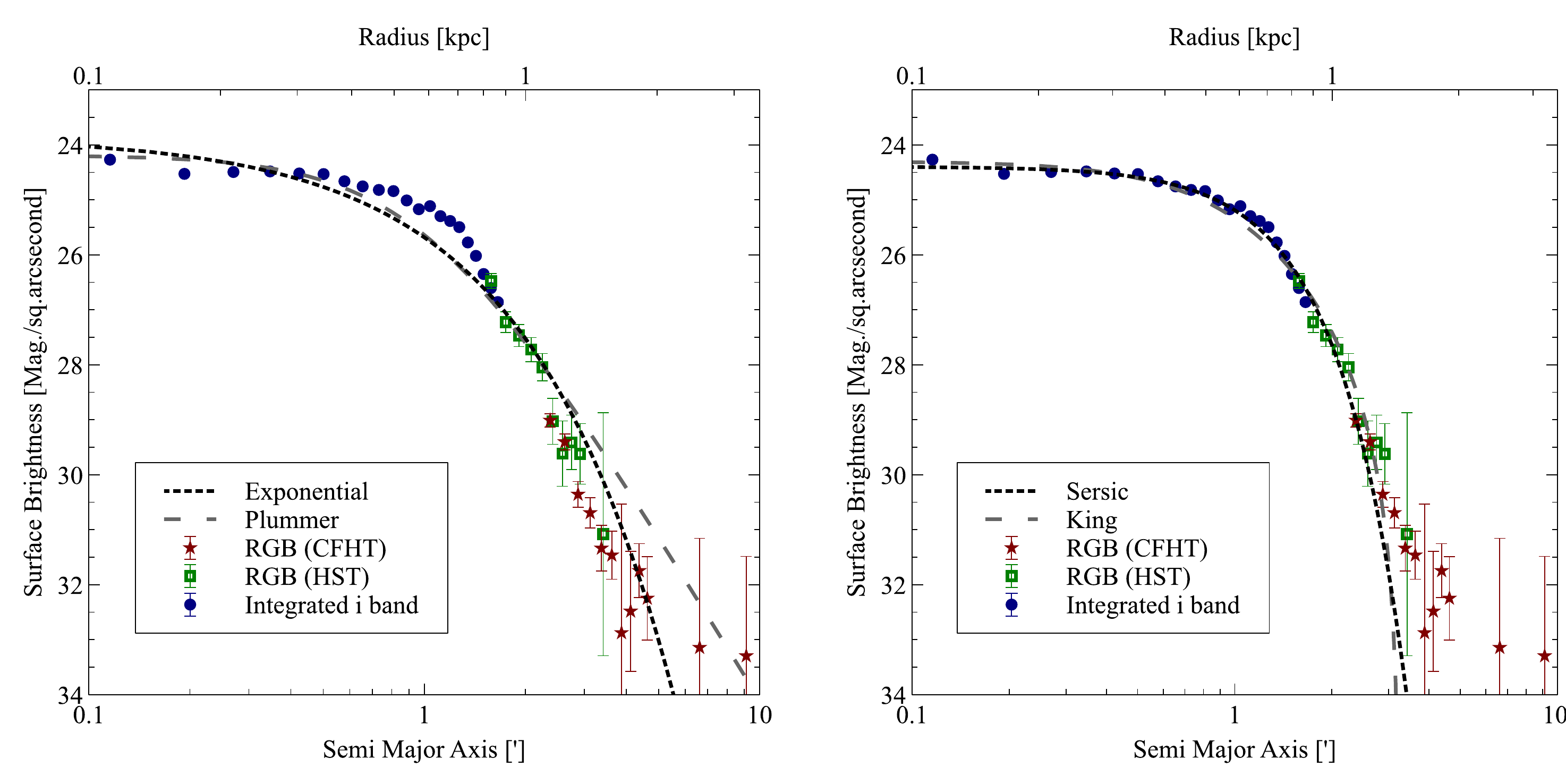}
\caption{Radial profile for Sag DIG, using a combination of integrated light (blue points) and star counts from HST (green squares) and CFHT (red stars). The star count profiles are scaled vertically to match the integrated light profile in overlapping regions. The best-fit Sersic model is shown with the dashed line. See text for details.} \label{radprofs}
\end{center}
\end{figure*}

Using the mean position angle and ellipticity, a series of elliptical annuli are used to determine a radial profile for the RGB density map shown in Figure \ref{rgbstruc}, with the background value that was previously estimated subtracted from the final profile. This profile is shown as red stars in Figure \ref{radprofs}. Crowding in the CFHT imaging is an significant concern at radii smaller than approximately 2 arcmins. Within this radius,  the radial structure is well mapped by the integrated light analysis conducted earlier. The RGB density profile (in units of stars/square arcmin) can be scaled to match the integrated light profile (in units of mags/sq.arcsec) by requiring that the average values of the points in the overlapping region between the integrated light and CFHT/MegaCam profile align. The blue points in Figure \ref{radprofs} show the $i_0$ band integrated light profile derived in Section \ref{integrated}. 

Since the overlap region in the radial profile includes some points for which crowding issues are a concern, we choose to augment the star count data with similar star count data from  the HST/ACS analysis by \cite{Momany2005}.  A DAOphot stellar catalogue was retrieved from the Hubble Legacy Archive, as described by \cite{Momany2005}. Observations were made using three filters, F475W, F606W, F81W. In the same manner as described above, a CMD was constructed using the bluest (F475W) and reddest (F814W) filters and a set of tramlines used to isolate the RGB. As the HST field of view is much smaller than that of CFHT MegaCam, there are no ``background" reference fields, and the image is dominated by Sag DIG stars. The spatial distribution of RGB stars identified from the CMD are then used to construct a radial profile. The HST star counts data is shown in Figure \ref{radprofs} as green squares.

By combining the $i_o$ band integrated light profile, the ground-based CFHT/MegaCam star counts data and the HST/ACS star counts data, Figure \ref{radprofs} shows the complete radial profile for Sag DIG over fully 8 magnitudes in surface brightness (a factor of  1500 in luminosity), extending down to as faint as nearly 33 mags/sq.arcsec. 
The shape of the annuli is fixed and is consistent across the full radial range of the combined profile. We remind the reader that the RGB and integrated light profiles are, in fact, potentially tracing different populations. The integrated light traces the luminosity-weighted stellar populations, and so is biased towards bright stars (i.e. the young stellar population that dominates the central regions) whereas the RGB profile traces intermediate and old stellar populations. We have used the $i_0$ integrated light profile, since it is likely a closer match to the population traced by the RGB than the bluer $g_0$ filter. From the absence of any significant ``kinks" in the profile at the points at which we transition from one tracer to another, however, it appears that the net effect of using the different tracers is small.

We fit multiple profiles to all the points defining the overall  radial profile in Figure \ref{radprofs}, shown with the black lines. An exponential and a Plummer profile (with 2 free parameters) are shown in the left hand panel and Sersic and King fits (with three free parameters) are shown in the right hand panel. The exponential form used is:

\begin{equation}
I(r)=I_{o,e} exp \left(- \frac{r}{r_e}\right)~. \label{expeqn}
\end{equation} 

\noindent The Plummer fit used is given by:

\begin{equation}
I(r)=I_{o,p} {r_e^2}{(r_e^2+r^2)^{-2}}~. \label{plumeqn}
\end{equation} 

\noindent The functional form for the Sersic fit (e.g. \cite{Graham2005}) used is:

\begin{equation}
I(r)=I_{o,s} exp \left( - b_{n} \left( \frac{r}{r_e}\right) ^{\frac{1}{n}} \right) \mbox{ where } b_n=1.9992n-0.3271~. \label{sersiceqn}
\end{equation} 

\noindent Finally, the King profile is given by:

\begin{equation}
I(r)=I_{o,k}  \left( {\left(1+\left(\frac{r}{r_c}\right)^2\right)^{-\frac{1}{2}}}-{\left(1+\left(\frac{r_t}{r_c}\right)^2\right)^{-\frac{1}{2}}}  \right)^2~. \label{kingeqn}
\end{equation} 

In fitting to the full profile,  the uncertainty on the integrated light portion is artificially increased relative to the star-count portions. This increase partially reflects the additional systematic errors in this profile due to contamination from flux from foreground stars, and it gives more equal weight to the inner and outer portions of the profile in determining the best overall fit. The best-fit parameters (and derived quantities) for all fits are given in Table \ref{spartable}. 

Adequate fits are obtained for both the Sersic and King profiles, whereas the exponential and Plummer profiles are unable to describe simultaneously the inner and outer portions of the radial profile. Specifically, the profile is considerably more cored than an exponential, as revealed by the Sersic $n=0.49$. This is typical of many dwarf galaxies (e.g., \citealt{Ferrarese2006}). It is also clear that the best fit profiles systematically underestimate the surface brightness of  Sag DIG in its extreme outskirts (i.e., beyond $\sim 4$ arcmins, or $\sim1.3$ kpc). In this region, the profile is still monotonically decreasing, implying the presence of an extremely faint, extended population of RGB stars, the origins of which will be discussed in the next section.

\begin{table}
\centering
\begin{minipage}{.9\linewidth} 
 \caption{Summary of parameters for Sersic, exponential, King and Plummer fits to the surface brightness profile. For explanation of the parameters, see Equations \ref{expeqn}, \ref{plumeqn}, \ref{sersiceqn} and \ref{kingeqn}. The integrated magnitude and central surface brightness are derived for the $i$ band. }\label{spartable}
 \small
\begin{tabular}{@{}lll@{}}
\hline
\hline
Parameter & Value  \\
\hline
\hline
$r_{e}$ (Exponential)  & $0.59 \pm0.03$' & $0.199\pm0.010$ kpc\\
$r_{e}$ (Plummer) & $1.02 \pm0.05$' & $0.344\pm0.017$ kpc \\
$r_c$ (King) & $1.11\pm0.06$' & $0.37\pm0.02$ kpc\\
$r_t$ (King) & $3.25 \pm0.07$' & $1.10\pm0.02$ kpc \\
 $r_s$ (Sersic) & $0.949 \pm0.012$' & $0.320\pm0.012$ kpc\\
 $n$ (Sersic) & $0.49\pm0.02$ & \\
 $i_{total}$ (Sersic) & $14.8\pm0.2$&mags.\\
 $i_{abs} $(Sersic) & $-10.6\pm0.2$&mags.\\ 
  $\mu_o$ (Sersic$^a$) & $24.40\pm0.03$ & mags./sq.arcsecond\\

\hline
\hline
\end{tabular}
\begin{footnotesize}
\hspace{0cm}{ $^{a}$ Computed in the $i$ band.} \\
\end{footnotesize}
\end{minipage}
\end{table}

\section{Discussion}

\subsection{On the content and structure of Sag DIG}

\begin{figure*} %remove stars for half page
%  \vspace{100pt}
\begin{center}
\includegraphics[width=\linewidth]{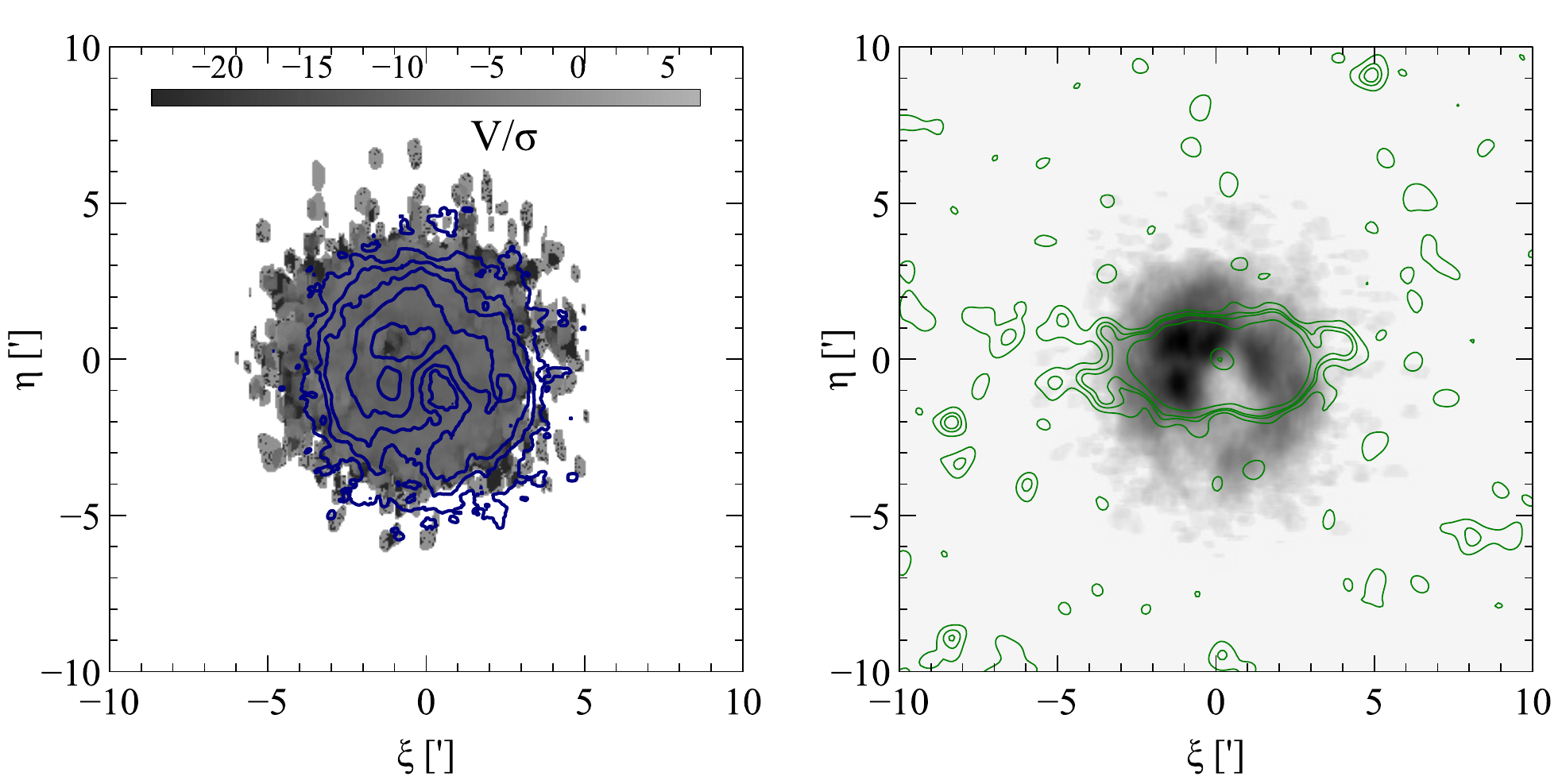}
\caption{\textit{Left:} HI is shown with contours at 10, 50, 100, 250, 500 (Jy/beam)(m/s) as seen previously in Figure \ref{FOV}. The underlying grey scale shows the velocity divided by dispersion from the analysis by \protect\cite{Hunter2012}. \textit{Right:} The RGB star distribution is shown in the contours (1.5, 3, 10 $\sigma$ above the mean background) overlaying the HI distribution in greyscale.}
\label{HI}
\end{center}
\end{figure*}

Our CFHT/MegaCam CMD of Sag DIG shown in Figure \ref{cmdfield} demonstrates the recent star formation and broad mix of stellar ages in this galaxy previously shown by multiple authors (e.g., \citealt{Karachentsev1999}, \citealt{Momany2005}, \citealt{Weisz2014}), and consistent with the findings for other isolated dwarf irregulars (see \citealt{Weisz2014}). Sag DIG contains a young population of massive blue stars (blue-loop and main sequence), the brightest of which imply star formation as recent as 50 Myrs years ago. A well populated RGB reveals a considerably older population, and allows a distance estimate using the TRGB of $(m-M)_0$=25.36$_{-0.14}^{+0.15}$ (corresponding to a distance of $1.18^{+0.08}_{-0.06} $ Mpc). The color of the RGB is remarkably blue, and if this effect is due to metallicity then it implies a mean metallicity of order [Fe/H]$ \simeq -2.2$\,dex. 
The blue edge of the RGB may have some contamination from the ``red" loop of He burning stars which could contribute to the RGB width as stars in these evolutionary phases overlap. However, if this contamination was significant we would expect to see clear evidence of a vertical sequence of red loop stars near the RGB, and none is present.
 Comparison of the inner and outer regions of Sag DIG shows that there is a spatial gradient in the stellar populations. The majority of the young stars are centrally concentrated, and the bright main sequence stars in the central regions appear to extend to younger ages than the bright main sequence stars at larger radii. The RGB stars in the central regions also appear to extend to bluer colors -- implying a larger contribution of RGB stars from intermediate age stellar populations compared to the older stars in the outer regions -- although this difference is not  statistically significant.

We can trace the low surface brightness structure of Sag DIG using the RGB star count maps. The main body of the dwarf is highly elliptical in shape ($e = 0.53$) with no obvious isophote twisting  as a function of radius. Instead, it looks similar to an inclined stellar disk (although we note that, in the absence of stellar velocities, it is also possible that Sag DIG is a highly ellipsoidal system). Under the assumption of an inclined disk, we can use the shape of the observed distribution of RGB stars to constrain the possible inclination angle of Sag DIG. As a low mass system, several observational studies have shown that we do not expect the stars to be in a thin disk akin to large galaxies (see \citealt{Roychowdhury2010, Sanchezjanssen2010}). From a theoretical perspective, we expect pressure support to be more important in low mass dark matter halos, leading to puffier disks \citep{Kaufmann2007}. We therefore assume that the galaxy is in a circular disk with a radius (A) equal to the semi-major axis, $a$, and a uniform thickness $C$. The observed semi-minor axis, $b$, is therefore a function of both $C$ and the inclination, $i$, such that 

\begin{equation}
b= Acos(i)+Csin(i)~.
\end{equation}

For the measured ellipticity of $e = 0.53$ and an infinitely thin disk,  then the minimum inclination of Sag DIG is $i \approx 62^{\circ}$. At the other limit -- i.e. assuming we are seeing Sag DIG edge-on -- then the maximum thickness of Sag DIG is $C/A \approx 0.48$. Thus, even if the main body of Sag DIG is a thickened disk, we are observing it relatively close to edge on.

As well as demonstrating Sag DIG to be a highly elliptical system, Figure \ref{rgbstruc} shows that the outermost contours of the galaxy (at $1.5\sigma$ and $3\sigma$ above the background) have a mild distortion that looks like a stretching of the contours along the major axis direction. It is difficult to interpret what, if anything, this feature indicates without ancillary data such as stellar radial velocities. If Sag DIG was closer to a large galaxy it could be interpreted tentatively as evidence for a tidal interaction. Additional insight into the extended structure of Sag DIG comes from analysis of the radial surface brightness profile. We augment the $i-$band integrated surface brightness profile of Sag DIG with RGB star counts from the CFHT/MegaCam data at large radii, and  RGB star counts from HST at intermediate radii. The resulting surface brightness profile extends over 8 magnitudes and is remarkably smooth. It is generally well fit by a Sersic or King profile over most of its radius (exponential and Plummer profiles with one less parameter do not do a satisfactory job of fitting the inner and outer regions of the profile simultaneously), but it still shows a very low level excess of stars at large radii ($> 3$ arcmins) in comparison to both of these model fits.

What could be the origin of this excess of stars at large radii in Sag DIG compared to the single component fits? \cite{Irwin1995} originally called this signature ``extra-tidal stars". These authors coined this phrase since they were fitting King profiles to Milky Way dwarf spheroidal satellites, and this excess was seen beyond the nominal King tidal radius. One physically plausible origin for these stars is that they were heated from the dwarf in its orbit around the Milky Way due to tidal effects. For Sag DIG, however, its extremely large distance from any large galaxy makes this unlikely, since tidal distortions are transitory in nature. Thus, even if Sag DIG had at one point been tidally harassed by a large galaxy, any signature would long since have disappeared (e.g. \citealt{Penarrubia2009}). If the very low level distortion on the outermost contours of the two dimensional surface brightness profile are related to this feature in the radial surface brightness profiles, then a more likely tidal scenario for Sag DIG is that it has experienced a merger of a much smaller object -- perhaps a globular cluster -- and that we are observing the remnants of this interaction. 

Another possible interpretation of the radial surface brightness profile, that we deem more likely than the tidal interaction scenario, is that Sag DIG is not well described by a single component model because Sag DIG does not consist of only a single component. Rather, there is a secondary, much more extended, lower surface brightness and lower luminosity component to the galaxy that surrounds the highly elliptical main body i.e., a stellar halo. Indeed, such a scenario is not particularly unusual; bright galaxies are known to possess multiple stellar components and more recent work on nearby dwarf galaxies suggests that multiple components are ubiquitous even at low stellar mass (e.g.,  \citealt{Martinezdelgado1999, Lee1999b, Lee1999c, Lee2000, Kleyna2003, Tolstoy2004, Kleyna2004, Vansevicius2004, Battaglia2006, Battaglia2008, Battaglia2011,Mcconnachie2007, Hidalgo2008, Bellazzini2014}). 
We note that the recent analysis of Sag DIG by \cite{Beccari2014} trace the profile to approximately 4 arcmin. and did not observe this excess. Comparison of the surface brightness profiles shows that our CFHT/MegaCam data and large field of view have allowed us to push to larger radii and lower effective surface brightness, which is required to identify this feature.

HI structure has been analyzed in detail previously by \cite{Lo1993}, \cite{Young1997} and \cite{Beccari2014} and it is interesting to compare this structure with our extended stellar maps. Figure \ref{HI} shows the integrated HI map of Sag DIG from the  LITTLE THINGS Survey \citep{Hunter2012} in comparison to the RGB stellar distribution discussed previously. Two aspects are particularly notable. First, the distribution of stars and gas are strikingly different within this galaxy, with the stars showing an elongated distribution and the gas showing a much more circular shape. We note that there is a prominent feature (a hole) in the HI distribution, and \cite{Momany2002} have previously suggested that the hole is due to stellar feedback, although they were unable to find any significant correlation between the location of the hole and the location of the most recent star formation for Sag DIG. Second, there is no clear signature of rotation of the gas in Sag DIG in the left panel of Figure \ref{HI}, showing $\frac{V_r}{\sigma}$. Indeed, the gas is either nearly entirely pressure supported or we are observing the disk face on. For the stars, we have established that if they are distributed in a disk, we must be observing this disk closer to edge on, not face on. It would be quite unusual to have a disk of gas presented nearly  perpendicular to the stellar disk, although not unprecedented, even in the Local Group. NGC6822 has been shown by \cite{Battinelli2006} to have the gas oriented at a large angle relative to the stars, making NGC6822 the closest example of a polar ring galaxy and requiring a merger at some point in its relatively recent past.

\subsection{Sag DIG in comparison to other Local Group dwarfs}

\begin{table}
 \centering
\begin{minipage}{0.9\linewidth} 
  \caption{Summary of parameters for both Sag DIG and Aquarius. }\label{galtable}

  \scriptsize
  \begin{tabular}{@{}llllll@{}}
  \hline
  \hline
      & Sag DIG &   Ref. &Aquarius &Ref. \\ 
\hline
 \hline
Distance [Mpc]                            & $1.16^{+0.08}_{-0.07} $ & (1)  &   $0.977\pm0.045 $ & (2)\\
E(B-V) [mag.]                               & 0.12$^{a}$  &  (3) & 0.045 & (3)\\
Stellar Mass [M$_{\odot}$]       &   $1.14\times10^6 $$^{b}$  & (1)       &  $1-2 \times 10^6$& (2,4,5) \\
HI Mass [M$_{\odot}$]               &  $8.8\times 10^6$   &   (6)    & $4.1\times 10^6$& (7) \\
Stellar [Fe/H] [dex]              & $\sim - 2.2$   &  (1) & $-1.44 \pm 0.03$ $^{c}$& (8)  \\
Mean Age [Gyrs.]                                  & 6  & (9)  & 13 & (9)   \\
Stellar Ellipticity                                     &  0.53 &  (1) & 0.6 $^{d}$ & (10) \\

\hline
\hline
\end{tabular}\\
\raggedright
\begin{footnotesize}
\hspace{0cm}{ References: (1) this work (2) \cite{Cole2014}  (3) \cite{Schlafly2011} (4) \cite{Young1997} (5) \cite{McConnachie2012} (6) \cite{Young1997} (7) \cite{Young2003}}  (8) \cite{Kirby2013} (9) \cite{Weisz2014} (10) \cite{McConnachie2006}\\
\hspace{0cm}{ $^{a}$ In agreement with Table 1, computed here as the mean value for each star, where as Table 1 is the central value.}\\
\hspace{0cm}{ $^{b}$ Computed using M/L$\sim 1$ and i$_{\odot}$ =4.58 mags  \cite{Blanton2003} }\\
\hspace{0cm}{ $^{c}$ Derived from spectra of individual RGB stars, with uncertainty reflecting the spread in metallicity between stars.}\\
\hspace{0cm}{ $^{d}$ In the central part, varies to 0.4 at larger radii.}\\
\end{footnotesize}
\end{minipage}
\end{table}

As an extremely isolated, very low mass dwarf galaxy, Sag DIG is an important target for understanding galaxy evolution at the lowest mass end. However, it is not unique. The closest known galaxy to Sag DIG is the Aquarius dwarf (DDO 210) at a separation  of $\sim 400$\,kpc. These two galaxies bear some similarities that make comparison of their properties particularly interesting. Both are located at large distances (\textgreater1 Mpc; \citealt{McConnachie2012}) from the Milky Way and M31, making recent tidal interactions with these large systems unlikely. Both are comparable in luminosity ($M_V = -11.5$ and $-10.6$ for Sag DIG and Aquarius, respectively), implying that their stellar masses are the same to within a factor of a few. Both systems are relatively gas rich, with HI masses of $M_{\ast}=4.1\times 10^6 M_\odot$ and $M_{\ast}=8.8\times 10^6 M_\odot$ for Aquarius and Sag DIG respectively.  The general properties of these two systems are summarized in Table \ref{galtable}.

Recently, the detailed star formation history of Aquarius has been determined through deep CMD analysis by \cite{Cole2014} that reaches one magnitude below the oldest main sequence turn-off. Aquarius is found to possess a very small old (\textgreater10 Gyrs) population, with a peak in its star formation approximately 6 - 8 Gyrs ago, after which the star formation rate has declined until the present day. Without similarly deep CMDs for Sag DIG, it is not possible to derive a SFH of similar quality for this galaxy. However, as previously discussed, the SFH derived by \cite{Weisz2014} suggests that Sag DIG is not dominated by an old population like many Local Group galaxies. On a qualitative note, the stellar content of Sag DIG and Aquarius are broadly similar; ground based CMDs for both galaxies (see \cite{McConnachie2006} for a ground-based CMD of Aquarius) have well defined RGB populations, weak AGB populations and significant numbers of bright main sequence/blue loop stars. However, in contrast to Sag DIG, Aquarius is not observed to be forming stars at the present, although it was forming stars as recently as a few tens of Myrs ago. For this reason, it is formally classed as a transition-type galaxy. At the low mass end, there may not be a fundamental distinction between a dwarf irregular and a transition-type galaxy, since low level star formation in a dwarf irregular may naturally lead to periods of little or no star formation (so-called ``gasping" e.g., \citealt{Tosi1991} and also \citealt{Weisz2011}). 

Both Aquarius and Sag DIG also share similarities in their HI structure relative to their stars. The main stellar bodies of both galaxies are relatively elliptical in extent yet both have integrated HI maps that imply a more spherical gas distribution (or a disk that is inclined at a significantly different angle to the stellar disk; see \cite{Young2003} for an HI map of Aquarius and \cite{McConnachie2006} for discussion of its structure relative to the stars). Sag DIG shows no evidence of rotation in the HI,  and Aquarius shows a low level systematic rotation with a peak velocity near $\sim$16 km s$^{-1}$ \citep{begum2004}. 

Can the similarities of these two isolated, low mass galaxies be used to imply common events in their evolution? Perhaps. The extended star formation of both Aquarius and Sag DIG is a feature common to all dwarf irregular/transition-type galaxies (e.g., \citealt{Weisz2011}). However, delayed star formation, such as that in Aquarius, appears to be relatively common for highly isolated low mass systems (see \citealt{Cole2014} for a discussion relating to Aquarius and Leo A) and so it will be important to determine if Sag DIG also follows this behaviour. Certainly, it is possible that the lack of any nearby large antagonists means that interactions at early times do not trigger star formation in these systems as might occur for closer satellites. Further, at the low mass end, the relationship between the accretion of dark matter and the accretion of baryons may be much more stochastic, such that these galaxies might not have accreted large amounts of gas at early times to fuel significant star formation. 
Reionisation is also often argued to have a significant impact on the early star formation of low mass dwarfs, but there is little evidence for this amongst those galaxies that have accurate star formations at early times (see \citealt{Weisz2014}). It is nevertheless conceivable that the gas that is accreted at early times in these low mass galaxies is heated by reionisation and does not cool sufficiently until later times at which point it start forming stars again.

Indeed, a process whereby the gas is heated in these small systems could be necessary to explain the difference between the stellar distribution (as traced by older RGB stars) and HI structures seen in Sag DIG and Aquarius.  In particular, it is plausible that the gas has been re-shaped by internal processes that heats up the gas, like star formation or supernova feedback. Once heated, the gas stays in the vicinity of the dwarf until it cools again. Once it has cooled again, the structure of the HI and the stellar structure of the galaxy can appear decoupled from each other, as is seen in these two galaxies. This idea is tentatively supported by the fact that the youngest stars in Aquarius are offset by several arcminutes from the center of the galaxy as traced by the oldest populations, and align with a slight ``dent" in the HI, perhaps suggesting that the gas is not virialised and the HI is relatively recently accreted. This scenario is also interesting as it applies specifically to low mass and isolated systems: for more massive dwarf galaxies, it is harder to heat the gas significantly, and so the gas and the stellar component are more likely to remain closely coupled; for satellite systems, once gas is heated and no longer at the center of the gravitational potential of the galaxy, it is more easily tidally stripped such that it does not have the chance to cool again. Clearly, however, Aquarius and Sag DIG are good examples of the inadequacy of assuming that the HI resides in a thin, rotationally supported disk, and that this disk bears any similarity to the structure  of the bulk of the stellar populations.

\section{Conclusion}

In this paper, we have introduced {\it Solo}, the {\it So}litary {\it Lo}cal Dwarfs survey. {\it Solo} is  a wide field photometric survey targeting every isolated dwarf galaxy within 3 Mpc of the Milky Way, based on $(u)gi$ multi-band imaging from CFHT/MegaCam in the north, and Magellan/Megacam in the south. All galaxies fainter than $M_V \simeq -18$ situated beyond the nominal virial radius of the Milky Way and M31 are included in this volume-limited sample, for a total of 42 targets.  We also present first results for  Sag DIG, one of the most isolated, low mass galaxies at the edge of the Local Group. 

The resolved stellar populations analyzed here are consistent with previous ground based studies of this galaxy. We provide updated estimates of its central surface brightness and integrated luminosity, and trace its surface brightness profile to a level fainter than 30 mags./sq.arcsec using a combination of integrated light and resolved stars. Sag DIG is well described by a highly elliptical (disk-like) system following a single component Sersic model, with a low-level distortion present at the outer edges of the galaxy. Were Sag DIG not so isolated, these distortions would likely be attributed to some kind of previous tidal interaction. We also show that Sag DIG has an excess of stars at large radii, over and above the number expected for a single component Sersic fit to the inner regions. We suggest that this is evidence of an extended stellar halo around this galaxy, although much deeper imaging data will be required to confirm this hypothesis to obtain sufficient star counts at large radii. 

The stellar and HI structures of SagDIG are compared with those in Aquarius, a similarly isolated low mass galaxy also at the edge of the Local Group. Both systems have a clear mismatch between their integrated stellar and HI structures. This difference in structures may be related to both their low mass and isolated locations, which permits the re-accretion of HI into their central regions after this gas was puffed up via some heating mechanism (likely earlier generations of star formation in the galaxy).

The derivation of a comprehensive set of parameters to describe the global properties of Sag DIG is the first step to the derivation of a comprehensive set of parameters for all galaxies in the {\it Solo} sample, using a homogeneous dataset and consistent analysis procedures. Our comparison of Sag DIG with Aquarius also  highlights the early utility of {\it Solo} in building up a more systematic understanding of the structures of dwarf galaxies in the very local Universe, that can inform interpretation of other systems for which resolved stellar analyses are not possible. The {\it Solo} dwarfs offer the opportunity to build a comprehensive understanding of the properties of low mass systems that have evolved in isolation for comparison to the satellite systems of the Local Group and other galaxy groups.
 
\section*{Acknowledgments}

Based on observations obtained with MegaPrime/MegaCam, a joint project of CFHT and CEA/DAPNIA, at the Canada-France-Hawaii Telescope (CFHT) which is operated by the National Research Council (NRC) of Canada, the Institut National des Science de l'Univers of the Centre National de la Recherche Scientifique (CNRS) of France, and the University of Hawaii. We thank the staff at CFHT for their help in obtaining the observations for the {\it Solo} sample.

Based on observations made with the NASA/ESA Hubble Space Telescope, and obtained from the Hubble Legacy Archive, which is a collaboration between the Space Telescope Science Institute (STScI/NASA), the Space Telescope European Coordinating Facility (ST-ECF/ESA) and the Canadian Astronomy Data Centre (CADC/NRC/CSA)

MGW is supported by National Science Foundation grants AST-1313045, AST-1412999.

GB gratefully acknowledges support through a Marie-Curie action Intra European Fellowship, (FP7/2007-2013; grant agreement number PIEF-GA-2010-274151), as well as the financial support by the Spanish Ministry of Economy and Competitiveness (MINECO) under the Ramon y Cajal Programme (RYC-2012-11537).

\nocite{*}
\bibliographystyle{apj}
\bibliography{bib_v7}{}

%\newpage
%\appendix
%\section[]{Sample Details}

\label{lastpage}

\end{document}